\documentclass[]{aa} %structabstract
\usepackage{graphicx}
\usepackage{txfonts}
\usepackage{natbib}
\bibpunct{(}{)}{;}{a}{}{,}

\begin{document}

\newcounter{subfigure}

\title{Properties of solar plage from a spatially coupled inversion of Hinode SP data}
\titlerunning{Properties of solar plage}

\author{D. Buehler \inst{1}, A. Lagg \inst{1},  S.K. Solanki \inst{1}$^,$\inst{2} \and M. van Noort \inst{1}}
\authorrunning{D. Buehler, et al.}

\institute{\inst{1} Max Planck Institute for Solar System Research, Justus-von-Liebig-Weg 3, 37077 G\"ottingen, Germany\\
\inst{2} School of Space Research, Kyung Hee University, Yongin, Gyeonggi, 446-701, Korea}

\date{}

\abstract
{}
{The properties of magnetic fields forming an extended plage region in AR 10953 were investigated.}
{Stokes spectra of the Fe I line pair at 6302 $\AA$ recorded by the spectropolarimeter aboard the Hinode satellite were inverted using the SPINOR code. The code performed a 2D spatially coupled inversion on the Stokes spectra, allowing the retrieval of gradients in optical depth within the atmosphere of each pixel, whilst accounting for the effects of the instrument's PSF. Consequently, no magnetic filling factor was needed.}
{The inversion results reveal that plage is composed of magnetic flux concentrations (MFCs) with typical field strengths of 1520 G at $\log(\tau)=-0.9$ and inclinations of $10^{\circ}-15^{\circ}$. The MFCs expand by forming magnetic canopies composed of weaker and more inclined magnetic fields. The expansion and average temperature stratification of isolated MFCs can be approximated well with an empirical plage thin flux tube model. The highest temperatures of MFCs are located at their edges in all $\log(\tau)$ layers. Whilst the plasma inside MFCs is nearly at rest, each is surrounded by a ring of downflows of on average 2.4 km/s at $\log(\tau)=0$ and peak velocities of up to 10 km/s, which are supersonic. The downflow ring of an MFC weakens and shifts outwards with height, tracing the MFC's expansion.  Such downflow rings often harbour magnetic patches of opposite polarity to that of the main MFC with typical field strengths below 300 G at $\log(\tau)=0$. These opposite polarity patches are situated beneath the canopy of their main MFC. We found evidence of a strong broadening of the Stokes profiles in MFCs and particularly in the downflow rings surrounding MFCs (expressed by a microturbulence in the inversion). This indicates the presence of strong unresolved velocities. Larger magnetic structures such as sunspots cause the field of nearby MFCs to be more inclined.}
{}


\maketitle

\section{Introduction}

In a typical active region on the solar disc three types of features can be identified most easily at visible wavelengths: sunspots, pores and plage. Whilst sunspots and pores are defined by their characteristic darkening of the continuum intensity, plage appears brighter than the surrounding quiet Sun mainly in spectral lines, or as faculae near the solar limb in the continuum. It has been known since the work of \citet{hale1908} that sunspots and pores harbour magnetic fields with strengths of the order of kG. \citet{babcock1955} showed that plage, too, is associated with magnetic fields, but it was only realised much later that it is also predominantly composed of kG magnetic features \citep{howard1972,frazier1972,stenflo1973}.\\
The kG magnetic fields, or magnetic flux concentrations (MFCs), in plage are often considered to take the form of small flux tubes or sheets, and considerable effort has gone into the details that determine their structure and dynamics \citep[see the review by][]{solanki1993}. The convective collapse mechanism \citep{parker1978,spruit1979,grossmann1998} is thought to concentrate the field on kG values \citep{nagata2008,danilovicft2010,requerey2014}, whereby the plasma inside the tube is evacuated and the magnetic field is concentrated. This mechanism may not always lead to kG fields, however \citep[e.g.][]{venkatakrishnan1986,solanki1996in,grossmann1998,socas2005}. The diameter of an individual kG flux tube is expected to be  a few 100 km or less, although a lower limit for kG fields may exist \citep{venkatakrishnan1986,solanki1996}. In the internetwork quiet Sun, diameters typically do not exceed 100 km, necessitating an instrument with an angular resolution of $0\farcs15$ or better to fully resolve an individual flux tube \citep{lagg2010}.\\ 
Owning to the comparatively small lateral size of these flux tubes, they are commonly treated using a thin flux tube model \citep{spruit1976,defouw1976}, where the lateral variation in the atmospheric parameters inside the tube is smaller than the pressure scale height. The 3D radiative MHD simulations by \citet{voegler2005} give rise to magnetic concentrations with properties that are close to those of the (2nd order) thin-tube approximation \citep{yelles2009}. More complex flux-tube models have also been proposed (see \citet{zayer1989} and references therein).\\
Despite their small size, many of the general properties of flux tubes residing in plage have nonetheless been determined by observations. This has been achieved by analyzing the polarisation of the light that is produced by the Zeeman effect in areas containing magnetic field \citep{solanki1993}. Thus, \citet{rabin1992,zayer1989} and \citet{rueedi1992} for example, used the deep photospheric infrared Fe I 1.56 $\mu$m line to find magnetic field strengths of $1400-1700$ G directly from the splitting of this strongly Zeeman sensitive line. Field strengths of around 1400 G were also obtained by \citet{wiehr1978} and \citet{martinez1997} when using lines in the visible, such as the 6302 $\AA$ line pair, whilst values of only $1000-1100$ G were found by \citet{stenflo1985} with the 5250 $\AA$ lines, which are formed somewhat higher in the photosphere. Finally, the Mg I $12.3$ $\mu$m lines, used by \citet{zirin1989}, returned values as low as $200-500$ G in plage regions, which are fully consistent with the kG fields observed in the aforementioned lines due to the even greater formation height of the Mg I $12.3$ $\mu$m lines, their different response to unresolved magnetic fields and the merging of neighbouring flux tubes \citep{bruls1995}.\\
The expansion with height of MFCs has also been investigated. \citet{pietarila2010expan} used SOT/SP images recorded at increasing $\mu$-values and examined the change in the Stokes $V$ signal of MFCs in the quiet Sun network. The variations of the Stokes $V$ signal across a MFC was, with the help of MHD simulations, found to be compatible with a thin flux tube approximation. \citet{martinezgonzales2012} analysed the Stokes $V$ area asymmetry across a large network patch recorded with IMaX \citep{martinez2011} aboard $\emph{\sc Sunrise}$ \citep{solanki2010,barthol2011} on the solar disc centre and found that the internal structure of the large network patch was likely to be more complex than that of a simple thin flux tube. A similar conclusion concerning the internal structure of plages was reached by \citet{berger2004}. \citet{rezaei2007} also examined the change of the Stokes $V$ area asymmetry of a network patch situated at disc centre using SOT/SP and showed that it was surrounded by a magnetic canopy. \citet{yelles2009} analysed thin flux tubes and sheets produced by MHD simulations and concluded that a 2nd order flux tube approximation is necessary to accurately describe the structure of the magnetic features. \citet{solanki1999} showed that the relative expansion of sunspot canopies is close to that of a thin flux tube, which could imply that the relative expansion of all flux tubes is similar.\\ 
The inclination with respect to the solar surface of MFCs in plage was found to be predominantly vertical, with typical inclinations of $10^\circ$ \citep{topka1992,martinez1997}, which can be attributed to the magnetic buoyancy of the flux tubes \citep{schuessler1986}, although MFCs with highly inclined magnetic fields were also found \citep{topka1992,bernasconi1995,martinez1997}. The azimuthal orientation of MFCs was shown to have no preferred direction \citep{martinez1997} and to form so called 'azimuth centres', although \citet{bernasconi1995} did find a preferred $E-W$ orientation.\\ 
The potential existence of mass motions inside MFCs has been fueled by the observation of significant asymmetries in the Stokes $Q$, $U$, and particularly $V$ profiles, in both amplitude and area \citep{solanki1984}. However, \citet{solanki1986} showed that within a MFC no stationary mass motions stronger than 300 m/s are present. This result was confirmed by \citet{martinez1997} using Milne-Eddington (ME) inversion results based on data of the Advanced Stokes Polarimeter (ASP) \citep{elmore1992}. Stokes profiles in plage display a marked asymmetry between the areas of their blue and red lobes \citep{stenflo1984}. This asymmetry is thought to result from the interplay between the magnetic element and the convecting plasma in which it is immersed \citep{grossmann1988,solanki1989}. Following this scenario \citet{briand1998} showed that low resolution Stokes $I$ and $V$ profiles of the Mg I $b_2$ line can be fitted with a combination of atmospheres representing a magnetic flux tube expanding with height, containing no significant flows, which is surrounded by strong downflows of up to 5 km/s representing the field-free convecting plasma around it. This scenario is generally supported by magneto-hydrodynamic (MHD) simulations performed by e.g. \citet{deinzer1984,grossmann1988,steiner1996,voegler2005}, although some of the simulated magnetic features do display internal downflows.\\
More recently, observations performed at higher spatial resolution using the Swedish Solar Telescope \citep[SST;][]{scharmer2003} have further confirmed the basic picture \citep{vandervoort2005}. \citet{langangen2007} found, by placing a slit across a plage-like feature, downflows in the range of $1-3$ km/s at the edges of the feature. Similarly, \citet{cho2010} observed, using SOT/SP, that pores, too, are surrounded by strong downflows in the photosphere.\\
The relationship between the magnetic field strength and continuum intensity of plage was studied extensively by \citet{kobel2011} using ME inversions of SOT/SP data and a clear dependence of the continuum intensity on the magnetic field strength was found \citep[c.f.][]{topka1997,lawrence1993}. Furthermore, the granular convection in plage areas has an abnormal appearance \citep[e.g.][]{title1989, narayan2010}. \citet{moriaga2008} and \citet{kobel2012} concluded that the high spatial density of the kG magnetic fields causes a suppression of the convection process.\\
As illustrated by the above papers, which are only a small sample of the rich literature on this topic, there has been significant progress in our knowledge of plage in the last two decades. Nonetheless, no comprehensive study of plage properties using inversions has been published since the work of \citet{martinez1997}, which was based on $1\arcsec$ resolution data from the Advanced Stokes Polarimeter (ASP). In the following sections we aim to both test and expand upon our knowledge of the typical characteristics associated with plage using the results provided by the recently developed and powerful spatially coupled inversion method \citep{vannoort2012} applied to Hinode SOT/SP observations. We concentrate here on the strong-field magnetic elements and do not discuss the horizontal weak-field features also found in active region plage areas \citep{ishikawa2008,ishikawa2009}.\\

\section{Data}

The data set used in this investigation was recorded by the spectropolarimeter \citep{lites2013}, which forms part of the solar optical telescope (SOT/SP) \citep{tsuneta2008sot,suematsu2008sot,ichimoto2008sot,shimizu2008sot} aboard the Hinode satellite \citep{kosugi2007}. The observation was performed on the $30^{th}$ of April 2007, UT 18:35:18 - 19:39:53, using the normal observation mode, hence, a total exposure time of 4.8 s per slit position and an angular resolution of $0\farcs3$ was achieved. All four Stokes parameters, $I$, $Q$, $U$ and $V$, were recorded at each slit position with a noise level of $1 \times 10^{-3} I_c$. The field of view contains a fully developed sunspot of the active region (AR) 10953 with an extended plage forming region trailing it. During the observation the spot was located in the southern hemisphere towards the east limb, $-190 X$, $-200 Y$, at $\mu=0.97$ ($\mu=cos(|\theta|)$, where $\theta$ is the heliocentric angle). A normalized continuum image of the investigated region used in the inversion is shown in Fig. \ref{cont}. The data were reduced using the standard \emph{sp\_prep} routine \citep{lites2013} from the solar software package.

\section{Inversion Method}  

\begin{figure}
   \centering
   \includegraphics[width=8cm]{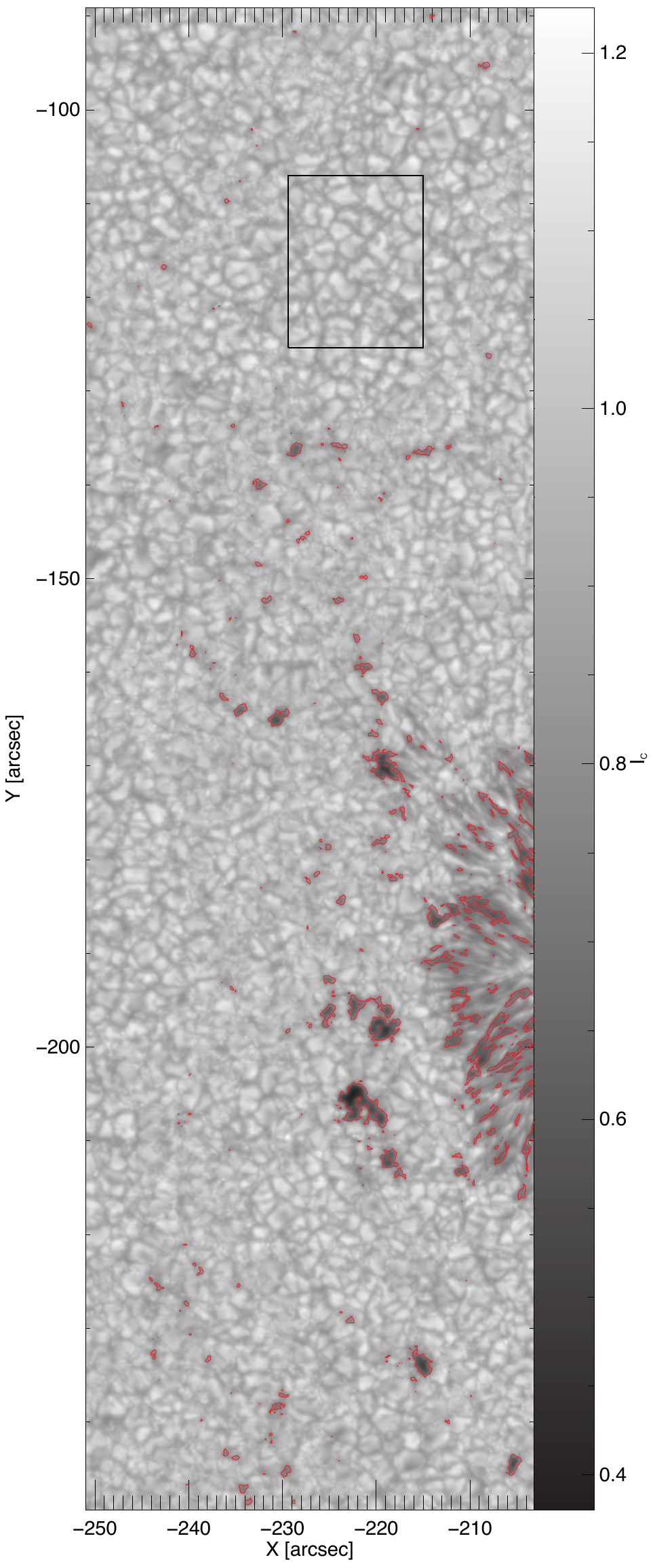}
      \caption{Normalized continuum intensity image of the region in which the Stokes profiles were inverted. Pixels where $T<5800$ K are enclosed by the red contour line. The $x$ and $y$ axes indicate the distance to the solar equator and central meridian, respectively. The $\emph{black}$ box denotes the area taken as a quiet Sun reference.}
         \label{cont}
   \end{figure}

The region of the SOT/SP scan containing most of the plage was inverted using the SPINOR code \citep{frutiger2000}, which uses response functions in order to perform a least-squares fitting of the Stokes spectra. It is based upon the STOPRO routines described by \citet{solanki1987}. The SPINOR code was extended by \citet{vannoort2012} to perform spatially coupled inversions using the point-spread-function (PSF) of SOT/SP. Such spatially coupled inversions have already been successfully applied to Hinode SOT/SP data of sunspots by \citet{riethmueller2013,vannoort2013,tiwari2013,lagg2014}. We employ the same PSF used by these authors, which is based on the work of \citet{danilovic2008}. The size of the inverted area, corresponding to that shown in Fig. \ref{cont}, is the largest that can currently be inverted in a single run by the employed code due to computer memory limitations. The inversion code allows the recovery of thermal, magnetic and velocity gradients with optical depth, among others, which reveal themselves by the strengths, shapes and asymmetries present in the SOT/SP Stokes profiles \citep{solanki1993,stenflo2010,viticchie2010}. The stratification of each atmospheric parameter with optical depth is calculated using a spline interpolation through preset $\log(\tau)$ nodes, where the code can modify a pixel's atmosphere. The resultant full atmosphere is then used to solve the radiative transfer equation and the emergent synthetic spectra are fitted iteratively by a Levenberg-Marquardt algorithm that minimizes the $\chi^2$ merit function.\\
The choice of the $\log(\tau)$ nodes is important for achieving a credible atmospheric stratification. Three nodes were chosen. A larger number of $\log(\tau)$ nodes produced more complex atmospheres at the expense of the uniqueness of the solution, while fewer $\log(\tau)$ nodes failed to fit the asymmetries in the observed spectra. The chosen nodes corresponded to optical depths at $\log(\tau)= 0, -0.9$ and $-2.3$ based on calculated contribution functions of the 630 nm line pair. The contribution functions were obtained from an empirical atmosphere simulating a plage pixel, i.e. containing magnetic field of 2000 G at $\log(\tau)=0$ and satisfying the thin-tube approximation at all heights. We also carried out inversions with nodes at slightly different $\log(\tau)$ values, to see if a better combination was available, but did not find one for plage. At each of the three chosen nodes the temperature, $T$, magnetic field strength, $B$, inclination relative to the line-of-sight (LOS), $\gamma$, azimuth, $\psi$, line-of-sight velocity, $v$, and micro turbulence, $\xi_{mic}$, were fitted, leading to 18 free parameters in total. We stress that no macro turbulent broadening was allowed and a fixed $\mu$-value of $\mu=0.97$ was assumed during the inversion. The influence of straylight from neighbouring pixels is taken into account by the PSF and the simultaneous coupled inversion of all pixels. Consequently, no magnetic filling factor was introduced in the inversion.\\ 
A common problem affecting this inversion process is the possibility that the fitting algorithm finds a solution that corresponds to a local $\chi^2$ minimum. This is particularly so if the initial guess atmosphere for a pixel is far from the global minimum. In an effort to ensure that the solution for each pixel of the inversion corresponds to the global minimum, the inversion process was performed a total of four times with each inversion performing 12 iterations. Save for the initial inversion each successive inversion used the smoothed results of the previous inversion as an initial input, thereby ensuring that the initial guess for each pixel is closer to the global minimum (under the assumption that the inversion does reach the global minimum by itself for the majority of pixels, but runs into danger of falling into a local minimum for a minority). After the fourth inversion process the mean $\chi^2$ value of all the pixels could not be decreased any further, e.g. by inverting the scan a successive time. 

\section{Results}
 
\begin{figure*}
       \centering
        \includegraphics[width=11cm]{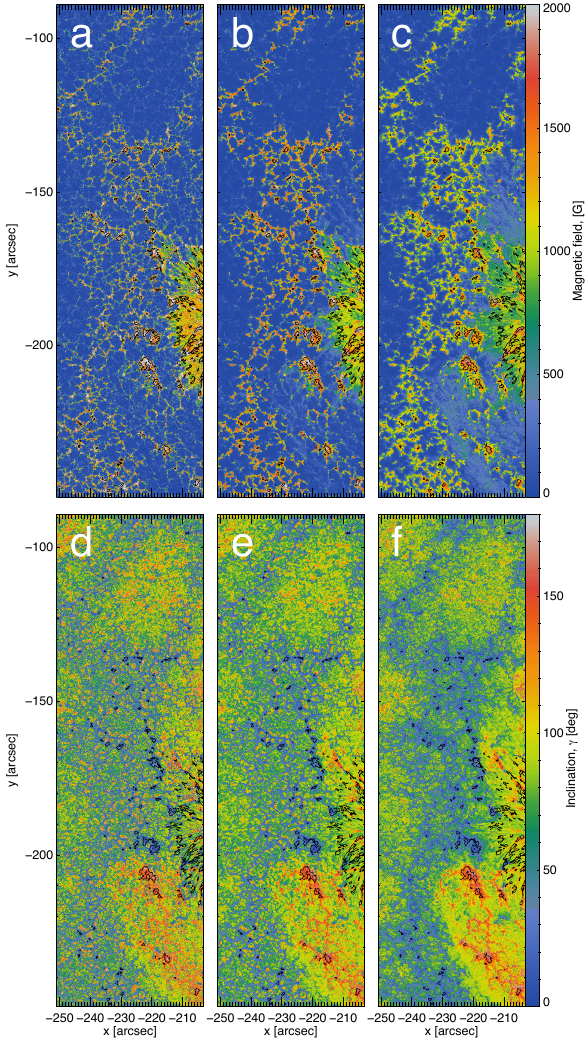}
        \caption{$\emph{a-c}$: Magnetic field strength retrieved by the inversion at $\log(\tau)=0, -0.9$ and $-2.3$, from left to right. The colour scale given on the right is identical in all the three images. $\emph{d-f}$: The line-of-sight inclination of the magnetic field obtained by the inversion at $\log(\tau)=0, -0.9$ and $-2.3$, from left to right. All three images have the same colour scale. The black contours in all images encompass pixels where $T<5800$ K.}
         \label{imgBIncl}
         \end{figure*}

\begin{figure}
   \centering
   \includegraphics[width=6cm]{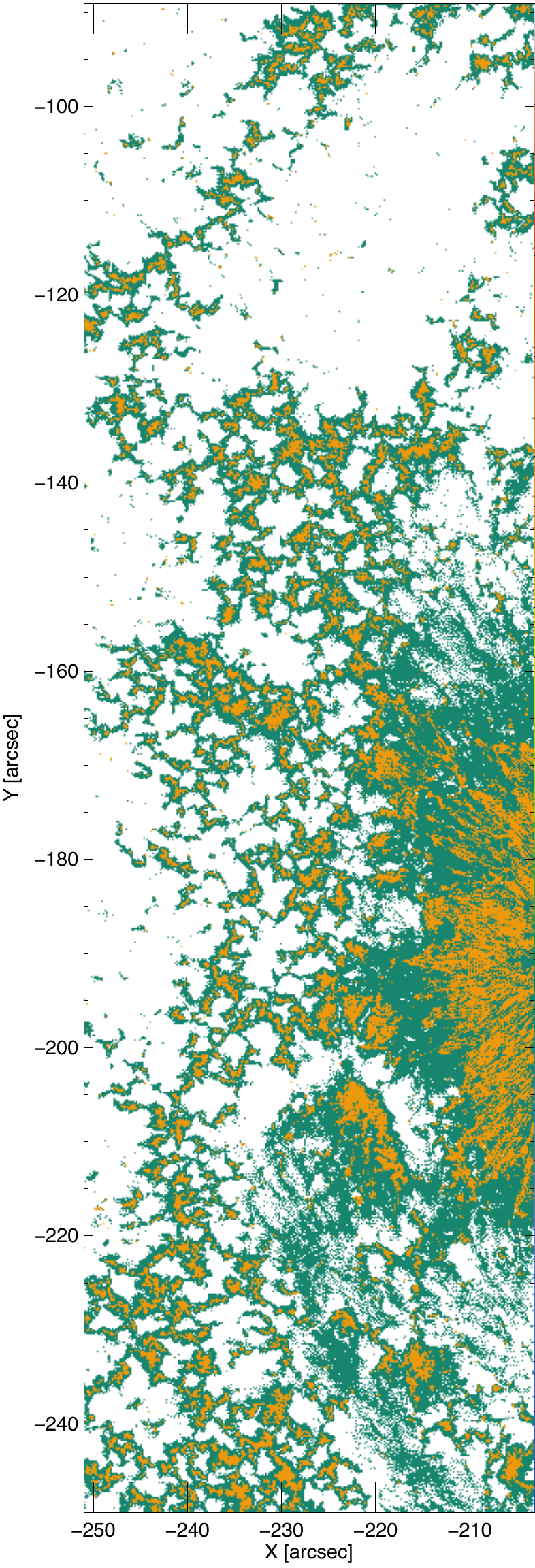}
      \caption{Image showing the location of $\emph{core}$ and $\emph{canopy}$ pixels using the definition given in Sect. 3. $\emph{Core}$ pixels are shown in orange and $\emph{canopy}$ pixels are coloured green. The white areas contain weak magnetic fields that were not considered to belong directly to the MFCs in plage regions.}
         \label{imgcanopy}
   \end{figure}

In this section we describe the various results obtained from the inversion. First we give a general overview of the output of the inversion followed by subsections dealing with more specific points. Figure \ref{cont} provides an overview of the continuum intensity, Figs. \ref{imgBIncl}a-c of the magnetic field strength returned by the inversion, and Figs. \ref{imgBIncl}d-f  of the inclination, $\gamma$, of the magnetic field vector. All these figures
display the entire field of view (FOV) to which the inversion code was applied. Fig. \ref{cont} reveals that part of the sunspot's penumbra as well as pores of various sizes are contained in the FOV and had to be excluded from the analysis. The sunspot's penumbra and the largest pores in the image were cut out by excluding the lower right hand side of the FOV from the analysis. However, many of the pores in the figure are only a few pixels in size, illustrated by the contour lines in Figs. \ref{cont} $\&$ \ref{imgBIncl}a-c and are often entirely embedded within a larger magnetic feature. These small pores were removed from the analysis using a temperature threshold of $T<5800$ K at $\log(\tau)=0$. Both higher and lower temperature thresholds, $\pm150$ K, were tested with insignificant effect upon the following results. The 5800 K threshold was finally chosen since it corresponds to the lowest temperature found in the quiet Sun, (see $\emph{black}$ box in Fig. \ref{cont}). An (alternative) intensity threshold to remove the pores yielded statistically similar results. The area within the $\emph{black}$ box serves as a quiet Sun reference throughout the following sections, since it is almost devoid of kG magnetic fields, although hG fields are present. The LOS velocities were corrected by subtracting an offset of 140 m/s, which was obtained by assuming that the pores are at rest on average.\\
Figures \ref{imgBIncl}a-c reveal that all MFCs expand significantly with height, suggesting that many pixels harbour magnetic fields only in higher layers of the atmosphere, indicating the presence of magnetic canopies. Therefore, all MFC pixels in the inversion result were subsequently divided into two populations: $\emph{core}$ pixels and $\emph{canopy}$ pixels. The $\emph{core}$ pixels were defined by a positive magnetic field gradient with optical depth, i.e. a magnetic field strength decreasing with height, and an absolute magnetic field strength, $B$, $>$ 1000 G at $\log(\tau)=0$. Higher thresholds at $\log(\tau)=0$ merely reduced the number of selected pixels but did not provide results that differ qualitatively from those presented here.\\
The $\emph{canopy}$ pixels were defined by a negative magnetic field gradient with optical depth and an absolute magnetic field strength, $B$, above 300 G at $\log(\tau)=-2.3$. A threshold $<$ 300 G at $\log(\tau)=-2.3$ caused the selection of a large number of pixels that were not directly connected to MFCs forming plage regions. These 'extra' pixels were predominantly associated with the sunspot penumbra and canopy, a filament and with weak horizontal magnetic fields found on top of granules in the few quiet Sun areas in Fig. \ref{cont}. This last group is likely related to the weak horizontal fields found in plage by \citet{ishikawa2009}. All selected pixels have a Stokes $Q$, $U$ or $V$ amplitude of at least 5$\sigma$, where $\sigma=1\times10^{-3}I_c$. The location of $\emph{core}$ and $\emph{canopy}$ pixels using these thresholds is illustrated in Fig. \ref{imgcanopy}. The figure shows that the $\emph{canopy}$ pixels associated with MFCs surround the $\emph{core}$ pixels. The sunspot's canopy forms an extended ring around it and photospheric loops between the sunspot and adjacent, opposite polarity pores with field strengths of up to 1000 G at the loop apex can be seen. Such a loop structure is located at approximately $-215X$, $-200Y$. The sunspot's canopy extends particularly far, as elongated finger-like structures at $-210X$, $-160Y$. These fingers are presumably very low lying loops connecting the spot to MFCs. The clear division between these various magnetic structures is only possible by applying the inversion code in 2D coupled mode, since the inversion requires no secondary atmosphere and/or a filling factor and the remaining single atmosphere is given the freedom to differentiate between $\emph{core}$ and $\emph{canopy}$ fields.\\
Most of the MFCs in Figs. \ref{imgBIncl}d-f have the same, positive, polarity (shown in blue). Only in the lower right hand corner of the field-of-view (FOV) can MFCs of negative polarity be found (shown in red in Figs. \ref{imgBIncl}d-f). The dominant polarity of the MFCs is opposite to that of the sunspot. Between the MFCs of opposite polarities a polarity inversion line (PIL) can be seen, stretching from approximately $-212X$, $-245Y$ to $-227X$, $-210Y$ in Figs. \ref{cont} \& \ref{imgBIncl}. H$\alpha$ images, not shown here, indicate the presence of a filament along this PIL. The photospheric part of this filament is visible in Fig. \ref{imgBIncl} at $\log(\tau)=-2.3$ in the form of predominantly horizontal magnetic fields (Fig. \ref{imgBIncl}f) of around 350 G. The atmosphere below the PIL is almost free of  magnetic field (Fig. \ref{imgBIncl}a). The location of the filament is clearly seen as the elongated canopy-like structure following the PIL in Fig. \ref{imgcanopy}. For a more detailed analysis of this filament the reader is referred to \citet{okamoto2008,okamoto2009}. Here we can add to their findings that, although the filament's magnetic field reaches down into the photosphere, it is largely restricted to layers more than roughly 200 km above the solar surface. This geometrical height was obtained from the hydrostatic atmospheres returned by SPINOR. The B value of 350 G in the filament is comparable (within a factor of 2) to the field strengths found in AR filaments by \citet{xu2010,kuckein2012,sasso2011} in the chromosphere sampled by the He I 10830 $\AA$ triplet. The azimuthal orientation of the magnetic field within the filament is almost invariant across the whole filament. Also, the orientation is not aligned with the sunspot's canopy, but rather almost parallel to the axis of the PIL, indicating sheared magnetic fields. This excludes the possibility that the field we assign to the filament could merely be a low-lying part of the sunspot's canopy.\\     
       \begin{figure*}
       \centering
        \includegraphics[width=18cm]{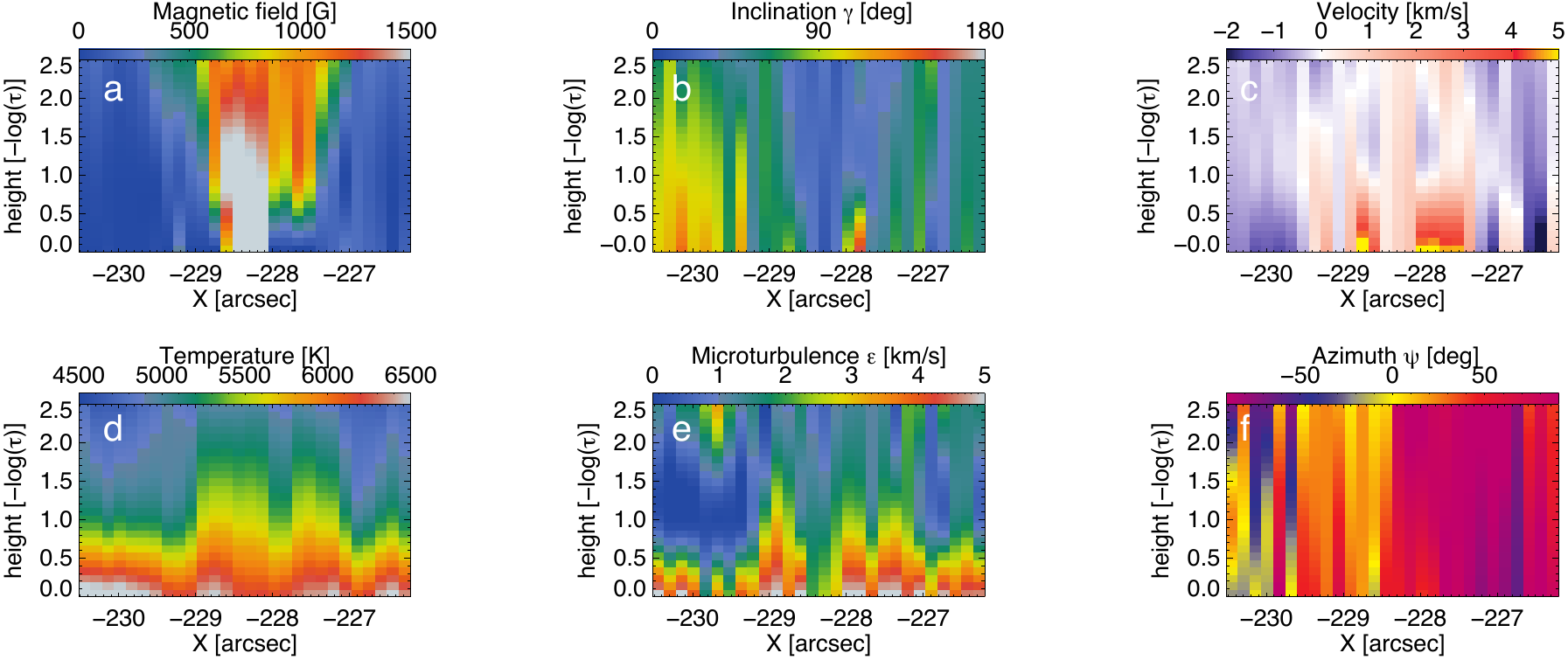}
        \caption{Vertical slice through a typical MFC. The Y coordinate of this MFC is -151". $\emph	          {a-c}$: Magnetic field, LOS inclination and LOS velocity from left to right.  $\emph{d-f}$: Temperature, microturbulence and azimuth from left to right.}
         \label{slice}
         \end{figure*}
Fig. \ref{slice} displays a vertical cut through a typical MFC and qualitatively illustrates many properties analysed in more detail in the following sections. The MFC is composed of nearly vertical kG magnetic fields that decrease with height, whilst the MFC expands. The apparent asymmetric expansion of the feature at $-288X$ arises from the merging of the feature's canopy with the canopy of a nearby MFC.  Both the temperature and the microturbulence are enhanced at mid-photospheric layers within the MFC, but even more so at the interface between it and the surrounding quiet Sun, where strong downflows are also present. The feature lies between two granules, which can be identified in the temperature image at $\log(\tau)=0$. The pixel-to-pixel variations seen in Fig. \ref{slice} are sizeable, but statistically the results are quite robust in that they apply to most plage MFCs.        
   
\subsection{Magnetic field strength}

\begin{figure}
   \centering
   \includegraphics[width=7cm]{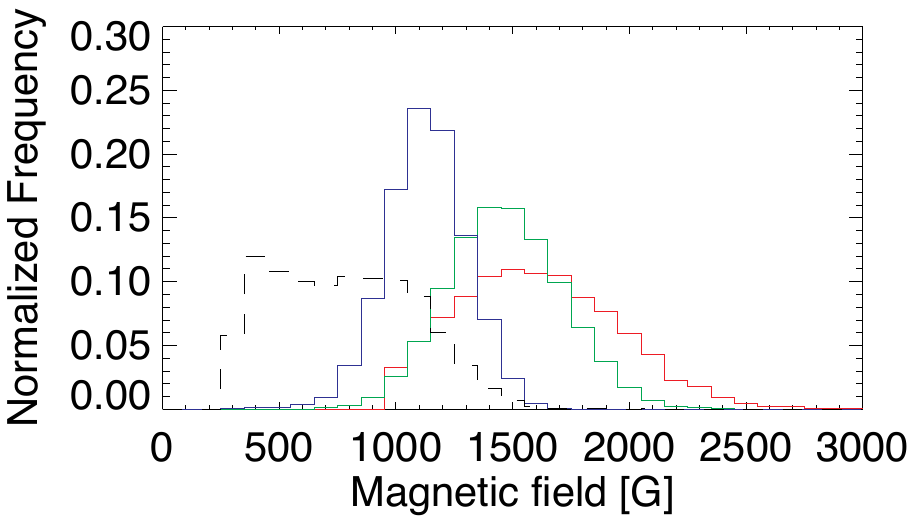}
      \caption{Histograms of $B$ values found in MFCs. The three coloured histograms are restricted to $\emph{core}$ pixels, where \emph{red} refers to $\log(\tau)=0$, \emph{green} indicates $\log(\tau)=-0.9$ and \emph{blue} refers to $\log(\tau)=-2.3$. The $\emph{dashed}$ histogram shows the field strengths of $\emph{canopy}$ pixels at $\log(\tau)=-2.3$.}
         \label{genB}
   \end{figure}

Figures \ref{imgBIncl}a-c indicate that the MFCs in plage regions are composed of magnetic fields on the order of kG in the lower and middle photosphere. This is confirmed by the histograms of magnetic field strength in Fig. \ref{genB}, which are restricted to pixels selected using the magnetic field thresholds defined in Sect. 4. Besides histograms of $B$ of $\emph{core}$ MFC fields at each optical depth, the histogram of the $\emph{canopy}$ pixels at $\log(\tau)=-2.3$ is plotted as well. Histograms of the magnetic field strength for the $\emph{canopy}$ at $\log(\tau)=0$ and $-0.9$ have been omitted as at these heights the atmosphere is similar to the quiet Sun or contains other, weaker fields that are analysed in Sect. 4.9. According to Fig. \ref{genB} the magnetic field strength at $\log(\tau)=-0.9$ has an average value of 1520 G. At this height the two Fe I absorption lines show the greatest response to all the fitted parameters, making the results from this node the most robust and comparable to results obtained from Milne-Eddington (ME) inversions \citep[e.g.][]{martinez1997} of this line pair. As expected, the average magnetic field strength in $\emph{core}$ pixels decreases with decreasing optical depth, so whilst at $\log(\tau)=0$ the average field strength is 1660 G, at $\log(\tau)=-2.3$ the average field strength drops to 1180 G. 
Fig. \ref{genB} also reveals that the widths of the histograms using $\emph{core}$ pixels decreases with height. At $\log(\tau)=0$ the FWHM of the histogram is 800 G that then subsequently decreases to 580 G at $\log(\tau)=-0.9$ and to 400 G $\log(\tau)=-2.3$, which is half the value measured at $\log(\tau)=0$. The comparatively broad distribution at $\log(\tau)=0$ appears to be intrinsic to the MFCs. Large MFCs display a magnetic field gradient across the feature, beginning at one kG at its border and rising to over two kG within the space of $\approx0\farcs5$. Smaller MFCs, too, often decompose into several smaller features when higher magnetic field thresholds are used. However, it cannot be completely ruled out, that the field strengths of the smallest MFCs are partially underestimated due to the finite resolution of SOT/SP. At greater heights neighbouring MFCs merge to create a more homogenous magnetic field with a smaller lateral gradient in B and appear to loose some of their underlying complexity. Nonetheless, differences in the distance between neighbouring MFCs still can lead to an inhomogeneous magnetic field strength above the merging height of the field \citep{bruls1995}, which itself strongly depends on this distance. \\
The distribution of the $\emph{canopy}$ pixels indicated by the $\emph{dashed}$ line in Fig. \ref{genB} reveals that $B$ in the $\emph{canopy}$ is generally much weaker than in the $\emph{core}$ pixels. The distribution also has no obvious \emph{cut-off} save for the arbitrary 300 G threshold, suggesting that the MFCs keep expanding with height in directions in which they are not hindered by neighbouring magnetic features.\\

\subsection{Magnetic field gradient}

\begin{figure}
   \centering
   \includegraphics[width=7cm]{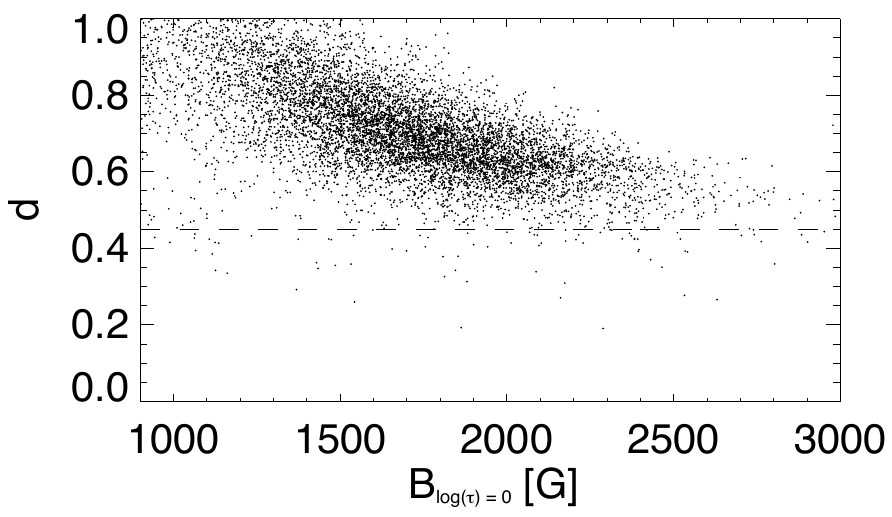}
      \caption{Ratio between field strengths at different $\log(\tau)$, $d$, calculated using Eq. 3, of $\emph{core}$ pixels plotted against the local solar coordinate corrected LOS magnetic field, $B$. The $\emph{dashed}$ line indicates the $d$ of a thin flux tube with $B$=2000 G at $\log(\tau)=0$.}
   \label{ftexpan_B}
   \end{figure}

The change in the peak magnetic field strength in each of the coloured histograms in Fig. \ref{genB} (see Sect. 4.1) indicates that the magnetic field strength of the $\emph{core}$ pixels decreases with height. This was investigated further, first, by correlating the relative decrease, $d$, of the magnetic field for each $\emph{core}$ pixel using   
	\begin{equation}
	d = \frac{B(\log(\tau)=-2.3)}{B(\log(\tau)=0)},
	\end{equation}
against $B(\log(\tau)=0)$, which is displayed in Fig. \ref{ftexpan_B}. The dashed line in Fig. \ref{ftexpan_B} shows the $d$ predicted by a thin flux-tube model with $B$=2000 G at $\log(\tau)=0$. The thin flux-tube model is identical to the plage model described by \citet{solanki1992pla}. \\
Fig. \ref{ftexpan_B} reveals that only magnetic fields above about 2000 G at $\log(\tau)=0$ have a $d$ close to the plage thin flux-tube model \citep{solanki1992pla}. The greatest limiting factor regarding the calculation of $d$ appears to suffer from a lack of magnetic flux conservation between the $\log(\tau)=0$ $\& -2.3$ nodes. The flux of isolated magnetic features at different $log(\tau)$ was calculated by drawing a box around it, which was large enough to easily encompass the entire MFC at all heights. Whilst the flux between the lower two nodes agreed within $5\%$, the $\log(\tau)=-2.3$ node consistently contained $20\%$ more flux than the $\log(\tau)=0$ node. This flux discrepancy between the nodes remained unchanged when the flux from a box containing several merged MFCs was used. Without this flux discrepancy, a $\emph{core}$ pixel with 2500 G at $\log(\tau)=0$ would have a $d$ value of 0.44 instead of 0.55, which would bring it close to the predicted value of the plage flux-tube model. Between the lower two $\log(\tau)$ nodes, where the flux is roughly conserved, the decrease in the magnetic field strength with height of such a pixel closely follows the thin flux-tube approximation. The $\emph{core}$ pixels with weakest $B$ are found at the edges of their respective MFC and thus may already be partially part of the canopy, due to the limited spatial resolution. This would reduce the vertical field strength gradient, thus increasing $d$ in particular for $\emph{core}$ pixels close to one kG. The small opposite polarity patches described in Sect. 4.9 allow the possibility of two opposite polarity magnetic fields to exist within a weak $\emph{core}$ pixel at the boundary of the MFC in the lower nodes. The Stokes $V$ signals from those two fields would at least partially cancel each other, leading to a reduction in the retrieved $B$ value (and apparent magnetic flux) in the lower two $\log(\tau)$ nodes and hence to an increase in $d$. Another contribution to the mismatch in Fig. \ref{ftexpan_B} is that some of the smallest $\emph{core}$ pixel patches are flux tubes which are not fully resolved by Hinode, in particular in the lower two layers. The expansion of such an unresolved flux tube would then take place primarily within the pixel, leading to a nearly unchanging $B$ in all three $\log(\tau)$ nodes, i.e. a $d$ close to unity.\\
The inversion also returns a geometric scale for each pixel. However, the inversion process only prescribes hydrostatic equilibrium within each pixel, but does not impose horizontal pressure balance across pixels. Therefore, each pixel has an individual geometric height scale,that can be off-set with respect to other pixels. This makes the comparison of gradients in (for example) $B$ between pixels with very different atmospheres difficult. However, the $\emph{core}$ pixels found at the very centre of a MFC, with $B$ > 2000 G, have very similar atmospheres, allowing the estimation of a common gradient in $B$. The gradient, $\Delta B$, in the magnetic field of these $\emph{core}$ pixels is $-2.6\pm0.5$ G/km between $\log(\tau)=0$ and $\log(\tau)=-2.3$.  As expected from Fig. \ref{ftexpan_B}, this $\Delta B$ is smaller than the gradient given by the thin flux tube model, which takes a value of $\Delta B=-3.9$ G/km over the same interval in $\log(\tau)$, but if magnetic flux conservation is imposed then the gradient of the inversion agrees with the thin-tube approximation.

\subsection{Expansion of magnetic features with height}
\begin{figure}
   \centering
   \includegraphics[width=7cm]{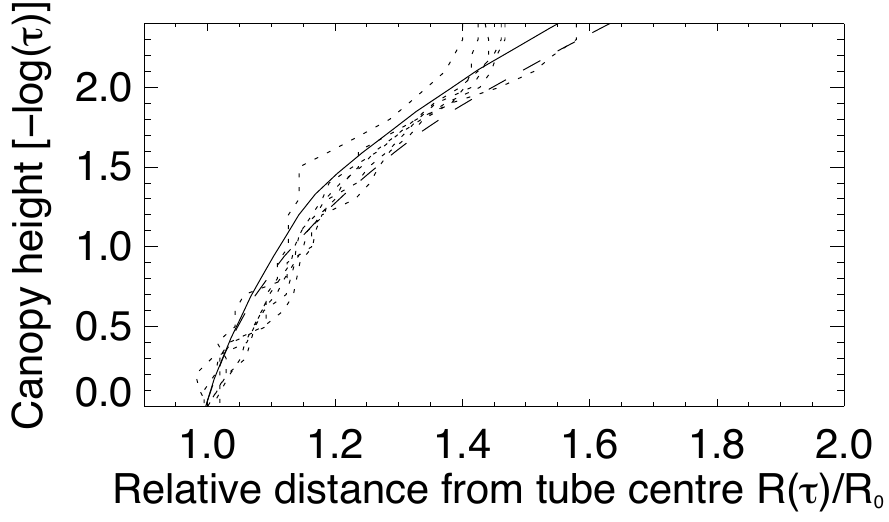}
      \caption{Relative expansion of five isolated magnetic features using Eq. 2 and shown by the $\emph{dotted}$ lines. The $\emph{solid}$ line represents the relative expansion of a zeroth order plage and the $\emph{dashed}$ line a network thin flux tube model.}
         \label{ftexpan_R}
   \end{figure}

Fig. \ref{imgBIncl} and, in particular, Fig. \ref{imgcanopy} demonstrate that the MFCs in the FOV expand with height. Furthermore, inclination and azimuth, reveal that MFCs generally expand in all directions and are not subjected to extreme foreshortening effects or deformations, (see Fig. \ref{genAziIncl} discussed in detail in Sect. 4.6), except due to other nearby MFCs (see Sect. 4.7). This raises the question of how close this expansion is to that of a model thin flux-tube. Several methods were tested to find a robust measure of the change in size of a magnetic feature with height. The most obvious method, the conservation of magnetic flux with height, was found to be unreliable to estimate the expansion of the magnetic features (see Sect. 4.2).\\
The expansion of the MFCs was, therefore, estimated in the following way. At $\log(\tau)=0$ the size of a magnetic feature was arbitrarily defined by the number of pixels that harboured a magnetic field of at least 900 G. Thresholds above and below this value ($\pm200$ G) did not significantly alter the results. Then all the magnetic field values in the $\log(\tau)=0$ image were normalized by the maximum field strength in the feature and the ratio, $r_t$, was calculated using $r_t=\frac{900 G}{B_{max}(\tau=1)}$. Each subsequent $\log(\tau)$ layer above $\log(\tau)=0$ was in turn normalized by its own $B_{max}(\tau)$ value. The expansion of a magnetic feature could then be tracked by the total number of pixels at a given $\log(\tau)$ layer where $\frac{B_(\tau)}{B_{max}(\tau)}>r_t$. This method assumes that the MFCs follow a self-similar structure with height. The thin-tube approximation as well as some other models \citep[e.g.][]{osherovich1983} follow this principle. Rather than tracking the expansion of a feature using only the three $\log(\tau)$ nodes returned by the inversion, the change of the magnetic field with height was tracked using a finer grid of $\log(\tau)$ layers, with a $\log(\tau)$ increment of 0.1. This finer $\log(\tau)$ grid was created using the same spline interpolation between the three nodes as was used during the fitting by the inversion procedure (see Fig. \ref{slice}). The 300 G threshold used to select $\emph{canopy}$ pixels in other parts of this investigation was not imposed here in order to avoid an artificial obstruction of the expansion.  Finally, the relative expansion of a feature was calculated using 
	\begin{equation}
	\frac{R(\tau)}{R_{0}} = \sqrt{\frac{A(\tau)}{A_{0}}}, 
	\end{equation}
where $R$ and $A$ are the radius and area, respectively, of the flux tube at optical depth $\tau$, $R_{0}$ and $A_{0}$ at $\log(\tau)=0$.\\
Five isolated magnetic features were selected from within the field of view. The number of selected features is small since most magnetic features have merged with other features at $\log(\tau)=-2.3$, as demonstrated in Figs. \ref{imgBIncl} \& \ref{imgcanopy}. The $R(\tau)/R_0$ of the five  selected features are represented in Fig. \ref{ftexpan_R} by the $\emph{dotted}$ curves, while
the $\emph{solid}$ line shows the relative radius of the $0^{th}$ order thin flux-tube plage model of \citet{solanki1992pla}. All the $\emph{dotted}$ lines in Fig. \ref{ftexpan_R} follow the expansion predicted by the plage model reasonably well. Interestingly, the thin flux-tube model for the solar network \citep{solanki1986}, \emph{dashed} line in Fig. \ref{ftexpan_R}, did not fit the expansion of the observed MFCs as well as the plage model ($\emph{solid}$ line in Fig. \ref{ftexpan_R}). The reduced relative expansion of the selected features above $\log(\tau)=-2$ when compared to the model may be an indication of the merging of features limiting the expansion at those heights. Another possibility is that a zeroth order model is not sufficient to describe the expansion of the selected features, especially in higher layers  \citep{yelles2009}, since the lateral variation of the magnetic field within the tube is no longer negligible in higher order flux tube models \citep{pneuman1986}. Given that the majority of MFCs merge with nearby MFCs, it follows that the majority of MFCs are expected to depart from the expansion displayed by isolated MFCs.

\subsection{Velocities}
    \begin{figure*}
   \centering
   \includegraphics[width=18cm]{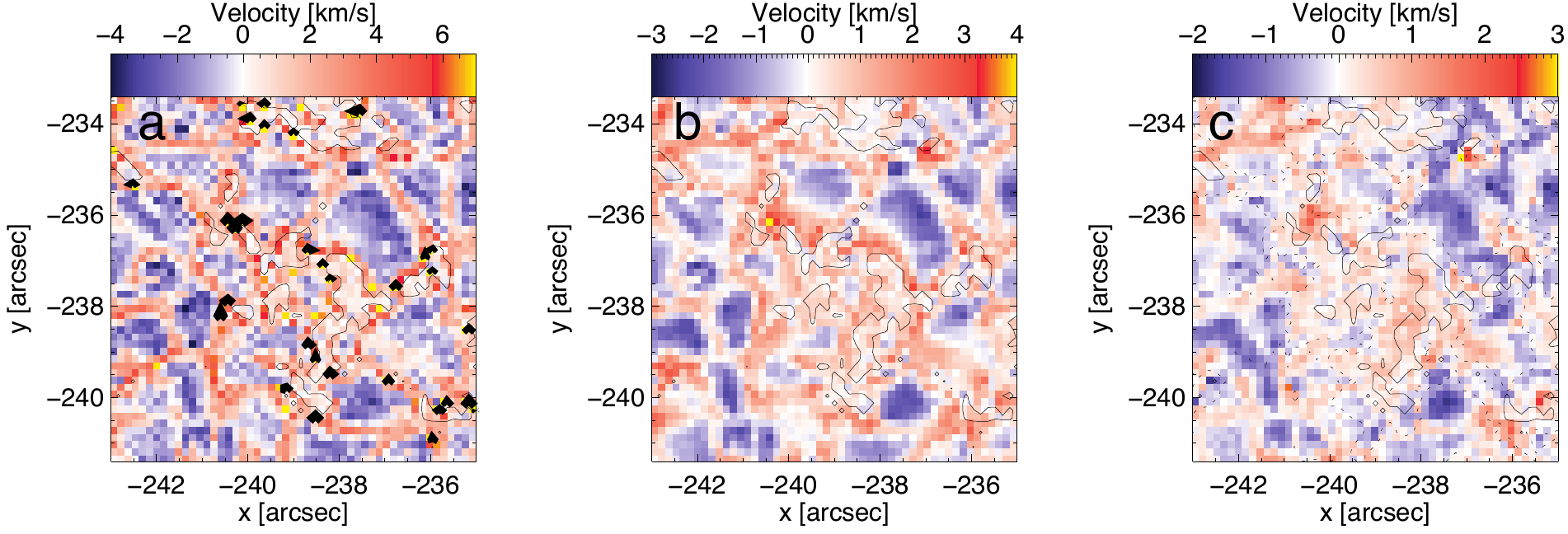}
   \caption{$\emph{a-c}$: Line-of-sight (LOS) velocities retrieved by the inversion at $\log(\tau)=0, -0.9$ and $-2.3$, from left to right. The $\emph{thin}$ black contours outline $\emph{core}$ pixels and the black areas at $\log(\tau)=0$ mark supersonic velocities. The $\emph{dotted}$ lines (in panel c) display $\emph{canopy}$ pixels.}
    \label{imgVel}
    \end{figure*}
\begin{figure}
   \centering
   \includegraphics[width=7cm]{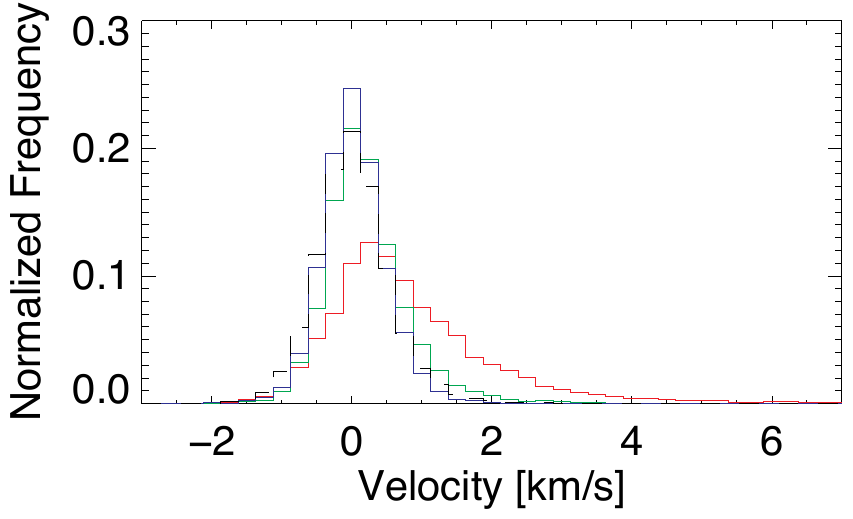}
      \caption{Histograms of the LOS velocities found in MFCs. The coloured histograms were obtained for $\emph{core}$ pixels, where \emph{red} refers to $\log(\tau)=0$, \emph{green} to $\log(\tau)=-0.9$ and \emph{blue} refers to $\log(\tau)=-2.3$. The $\emph{dashed}$ histogram shows the \emph{los} velocities of $\emph{canopy}$ pixels at $\log(\tau)=-2.3$.}
   \label{genVel}
   \end{figure}

\begin{figure*}
   \centering
   \includegraphics[width=18cm]{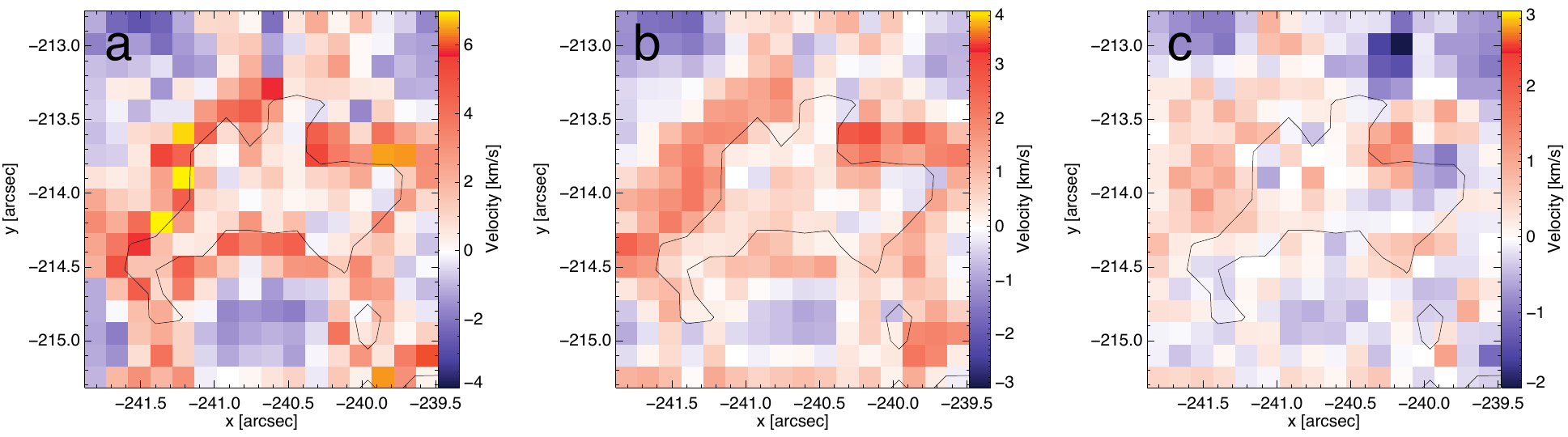}
   \caption{Same as Fig. \ref{imgVel}, but for a single MFC.}
    \label{imgVelzoom}
    \end{figure*}

\begin{figure}
   \centering
   \includegraphics[width=7cm]{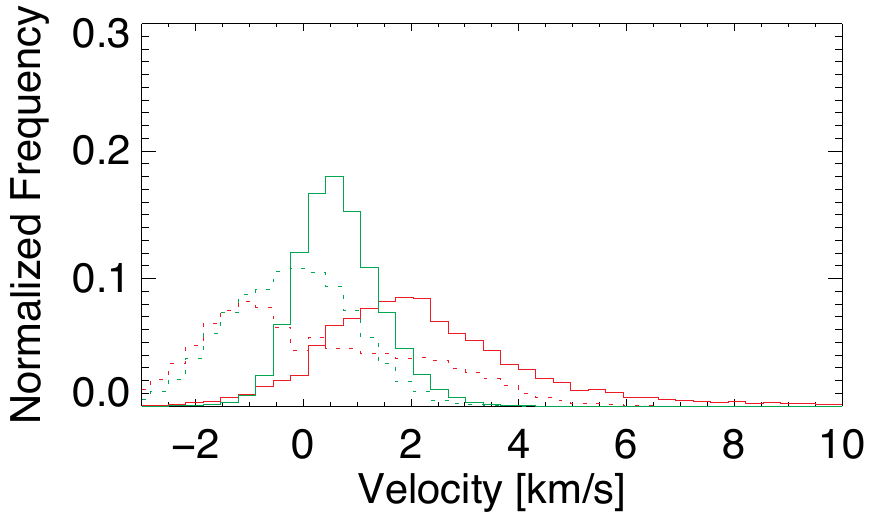}
      \caption{Histograms of the LOS velocities of pixels immediately surrounding $\emph{core}$ pixels, where \emph{red} refers to $\log(\tau)=0$, \emph{green} to $\log(\tau)=-0.9$. The $\emph{dotted}$ histograms display LOS velocities in the quiet Sun.}
         \label{Velring}
   \end{figure}

\begin{figure*}
   \centering
   \includegraphics[width=18cm]{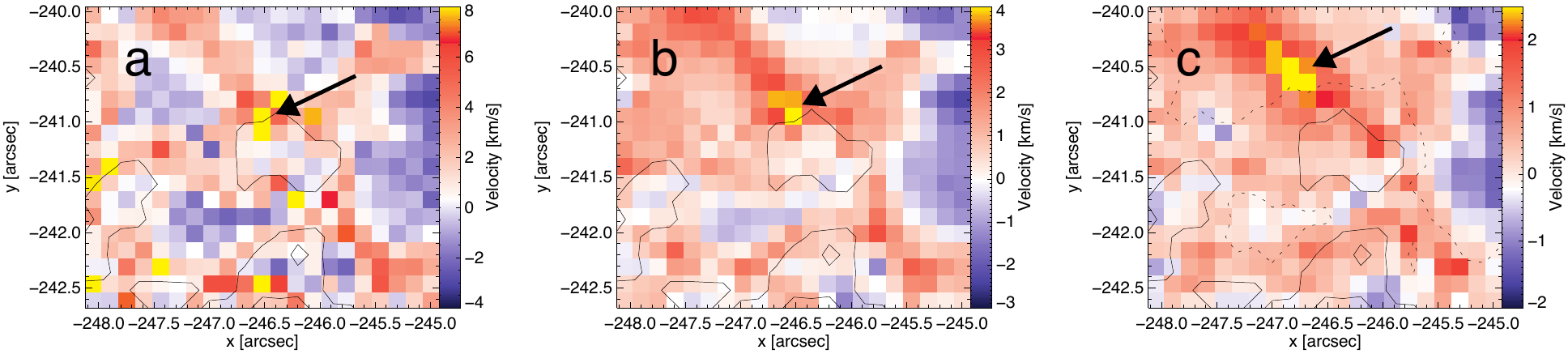}
   \caption{Same as Fig. \ref{imgVel}, but for a small part of the full FOV, chosen to reveal a downflow plume around a MFC. The $\emph{solid}$ contour line bounds $\emph{core}$ pixels and the $\emph{dotted}$ line (in panel c) $\emph{canopy}$ pixels. The arrows point to the location of the plume at each height.}
    \label{imgVelplume}
    \end{figure*}

  
Figure \ref{imgVel}a-c displays the LOS velocities retrieved by the inversion at the three $\log(\tau)$ nodes. The small FOV for this figure, representing a typical plage region, was chosen to better highlight the striking differences between velocities in field-free and kG regions and the unusual velocities recorded at the interface between these two regions. The $\emph{core}$ pixels in the images are enclosed by the $\emph{thin}$ black contour lines. Outside the areas harbouring $\emph{core}$ pixels, the typical granular convection pattern can be seen at $\log(\tau)=0$ $\&$ $-0.9$. The LOS velocities in the top node outline only traces of the stronger granules and display some similarities with chromospheric observations, albeit with smaller velocity amplitudes. The $\emph{dotted}$ contour lines outline the $\emph{canopy}$ fields.\\
The plasma within the majority of $\emph{core}$ pixels is nearly at rest in all $\log(\tau)$ nodes, as Figs. \ref{imgVel}a-c qualitatively indicate. Histograms of the LOS velocities of these pixels displayed in Fig. \ref{genVel} support this assertion. The mean and median velocities of these $\emph{core}$ pixels are 0.8 km/s and 0.6 km/s, respectively, at $\log(\tau)=0$. They drop to 0.2 km/s and 0.2 km/s at $\log(\tau)=-0.9$, and to 0.1 km/s and 0.0 km/s at $\log(\tau)=-2.3$.
The $\emph{red}$ histogram in Fig. \ref{genVel}, corresponding to the $\log(\tau)=0$ layer, also contains a significant fraction of pixels with downflows larger than 1 km/s. The tail of faster downflows is mainly responsible for the larger than average velocity at $\log(\tau)=0$. The individual MFCs, one of which is displayed in Fig. \ref{imgVelzoom}, were inspected further to determine the location and nature of these fast downflows at $\log{\tau}=0$. Fig. \ref{imgVelzoom}a reveals the $\emph{core}$ pixels of a MFC, enclosed by the black contour line, to be surrounded by a ring of strong downflows. Other MFCs show similar downflow rings at $\log(\tau)=0$ and can often be identified based on such a ring alone. Downflows that exceed 1 km/s are never found within a MFC, but occasionally a $\emph{core}$ pixel located at the edge of a MFC can coincide with a strong downflow, giving rise to the tail of strong downflows seen in the $\emph{red}$ histogram in Fig. \ref{genVel}. Furthermore, the velocities within the rings are not of uniform magnitude. Often some portions of a ring show very fast downflows, up to 10 km/s, whilst others can harbour flows of barely 1 km/s. The portions featuring the fastest downflows do not have a positional preference with respect to its MFC, of which Fig. \ref{imgVelzoom} is one example, which precludes that the magnitude of the downflow is affected by the viewing geometry. The downflow ring seen at $\log(\tau)=0$ in Fig. \ref{imgVelzoom}a is still visible at $\log(\tau)=-0.9$, in Fig.  \ref{imgVelzoom}b. However, the pixels with the fastest downflows in the ring seem to be located further away from the $\emph{core}$ pixels in this layer, when compared to Fig. \ref{imgVelzoom}a. It appears that the downflow ring shifts outwards as the magnetic field of the MFC expands with height (see Sect. 4.3). Also, the ring appears to be wider at this height. At $\log(\tau)=-2.3$ the ring can no longer be identified.\\ 
A quantitative picture of the LOS velocities within these rings was gained by analyzing the pixels, which directly adjoin the $\emph{core}$ pixels. Only the lower two layers were analysed and are displayed in Fig. \ref{Velring}, since the rings are no longer present at $\log(\tau)=-2.3$. LOS velocities of up to 10 km/s were found within these rings at $\log(\tau)=0$ and the mean and median values for the corresponding histogram are 2.44 km/s  and 2.16 km/s, respectively. For comparison, the fastest downflow in the quiet Sun region, see the black box in Fig. \ref{cont}, is only 6 km/s. At $\log(\tau)=-0.9$ the rings no longer contain downflow velocities faster than those found in the quiet Sun at the same $\log{\tau}$ layer, but the histogram of the velocities in the ring still has a mean LOS velocity of 0.84 km/s and a median of 0.77 km/s. We further investigated whether the fast downflow velocities at $\log(\tau)=0$ in these rings attain supersonic values. Since the SPINOR code calculates a full atmosphere, including density and pressure, for each pixel, we were able to directly calculate a local sound speed for each pixel over its entire $\log(\tau)$ range, using the same approach as \citet{lagg2014}. The adiabatic index, also needed to calculate the sound speed, was acquired from look-up tables produced by the MURaM MHD simulation code \citep{voegler2005}. Supersonic velocities were found in pixels with downflows exceeding 8 km/s at $\log(\tau)=0$ and the fastest downflows, reaching 10 km/s, have a Mach number of 1.25. Higher $\log(\tau)$ layers did not show any supersonic velocities in any of the pixels. Since the highest speed found in the quiet Sun reach up to 6 km/s, no supersonic velocities were consequently found in the quiet Sun. Furthermore, we determined that $2.5\%$ of a MFC's downflow ring contain supersonic velocities. Pixels containing supersonic velocities are coloured black in Fig. \ref{imgVel}.\\
Whilst the downflow rings seen at $\log(\tau)=0$ $\& -0.9$ are generally not traceable at $\log(\tau)=-2.3$, many MFCs additionally have downflows in the form of a plume-like feature, which can be traced through all three $log(\tau)$ layers. An example of such a plume-like feature is displayed in Fig. \ref{imgVelplume}. The plume lies just outside the $\emph{core}$ pixels at $\log(\tau)=0$ (Fig. \ref{imgVelplume}a), but is considerably further away, at the boundary of the canopy at $\log(\tau)=-2.3$ (Fig. \ref{imgVelplume}c), and appears to trace the expansion of the MFC. It also increases in size with height. At all heights the LOS velocities of the feature are high when compared to their immediate surroundings, but only at $\log(\tau)=0$ are the velocities in the feature higher than can be found in the quiet Sun. As with the downflow rings the velocities progressively increase with depth.\\    

\subsection{Temperature}
\begin{figure*}
   \centering
   \includegraphics[width=18cm]{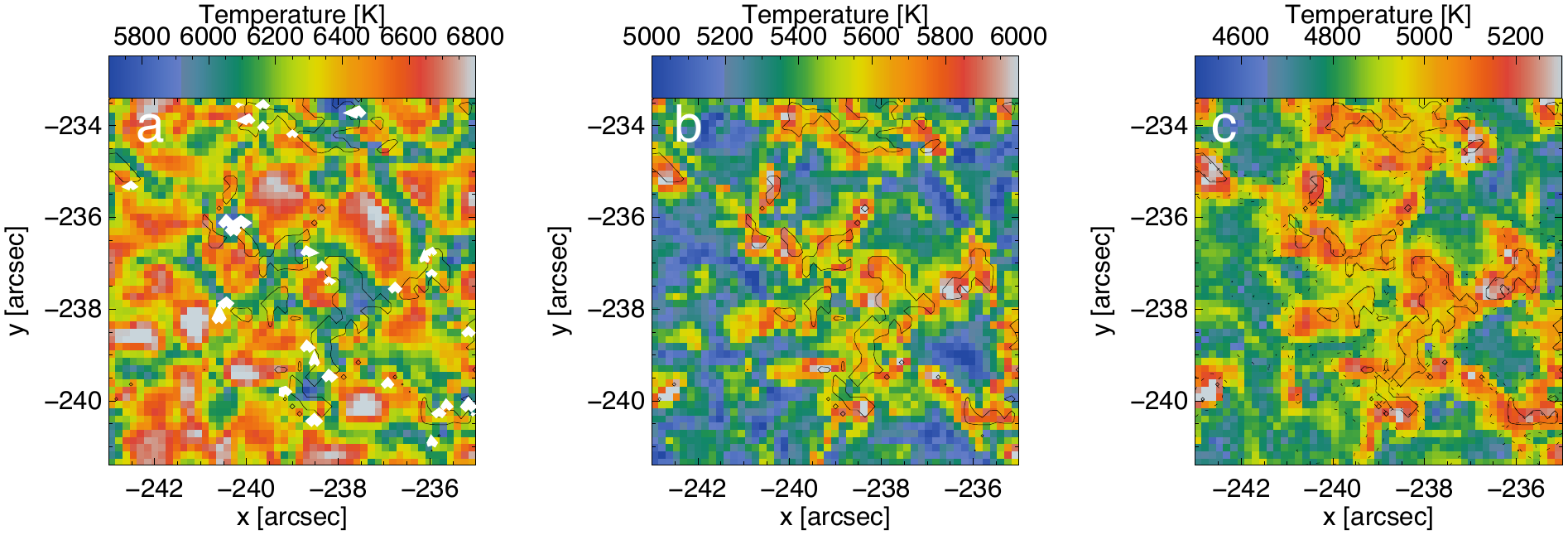}
   \caption{$\emph{a-c}$: Images of the temperature at $\log(\tau)=0, -0.9$ and $-2.3$, respectively. The black contour lines encompass $\emph{core}$ pixels and the white areas at $log(\tau)=0$ denote pixels containing also supersonic velocities. The $\emph{dotted}$ lines (in panel c) display $\emph{canopy}$ pixels.}
    \label{imgT}
    \end{figure*}

\begin{figure*}
   \centering
   \begin{minipage}{0.5\linewidth}
   \centering
   \includegraphics[width=7cm]{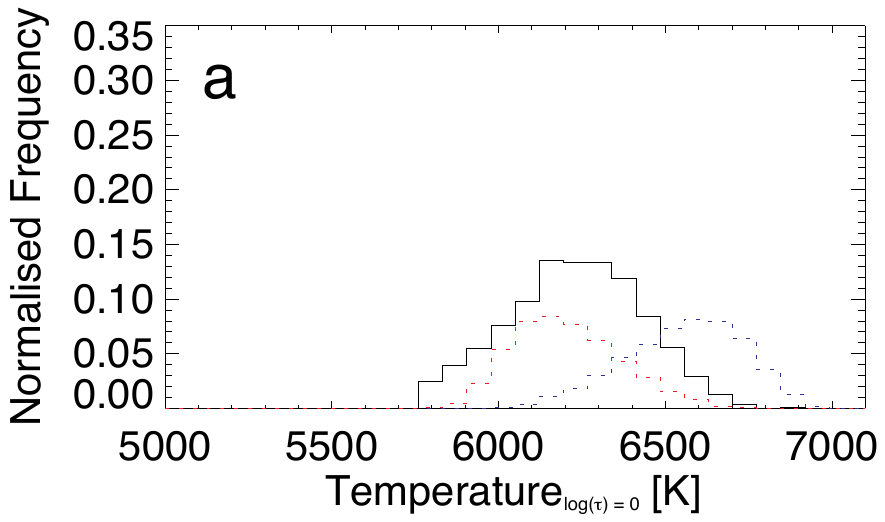}
   \end{minipage}%
   \begin{minipage}{0.5\linewidth}
   \centering
   \includegraphics[width=7cm]{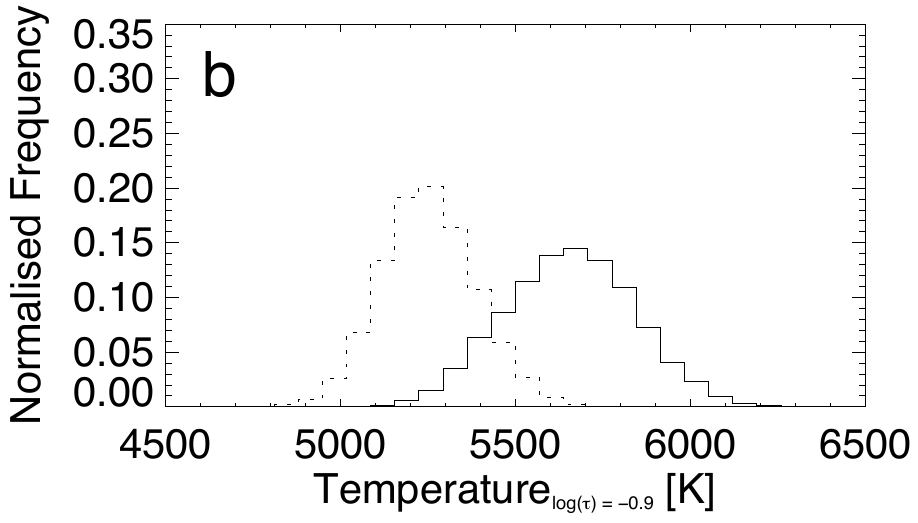}
   \end{minipage}
   \begin{minipage}{0.5\linewidth}
   \centering
   \includegraphics[width=7cm]{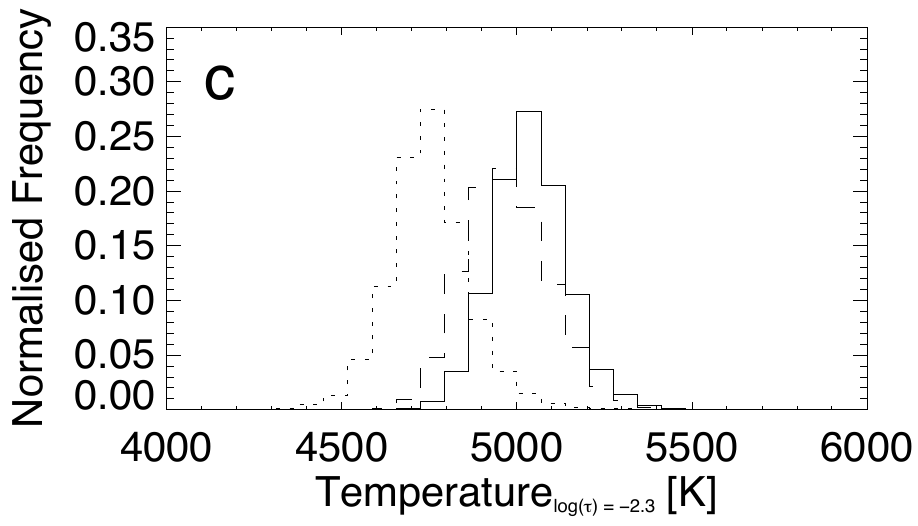}
   \end{minipage}% 
   \caption{$\emph{a}$: Temperature histograms at $\log(\tau)=0$ of $\emph{core}$ plage pixels, $\emph{solid}$ line, and the quiet Sun (\emph{dotted} lines). The quiet Sun temperatures have been further divided into temperature histograms corresponding to downflowing, $\emph{dotted red}$, and upflowing, $\emph{dotted blue}$, regions. $\emph{b}$: Temperature histograms at $\log(\tau)=-0.9$ of $\emph{core}$ plage pixels, $\emph{solid}$, and the quiet Sun, $\emph{dotted}$. $\emph{c}$: Temperature histograms  at $\log(\tau)=-2.3$ of $\emph{core}$ plage pixels, $\emph{solid}$, and the quiet Sun, $\emph{dotted}$. The $\emph{dashed}$ histogram displays the temperatures in $\emph{canopy}$ pixels.}
   \label{T}
   \end{figure*}

\begin{figure*}
   \centering
   \begin{minipage}{0.5\linewidth}
   \centering
   \includegraphics[width=7cm]{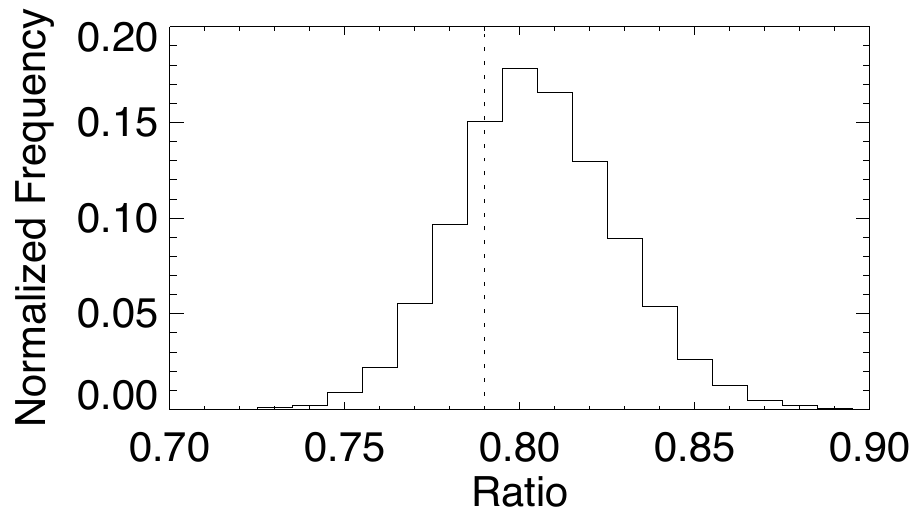}
   \end{minipage}%
   \begin{minipage}{0.5\linewidth}
   \centering
   \includegraphics[width=7cm]{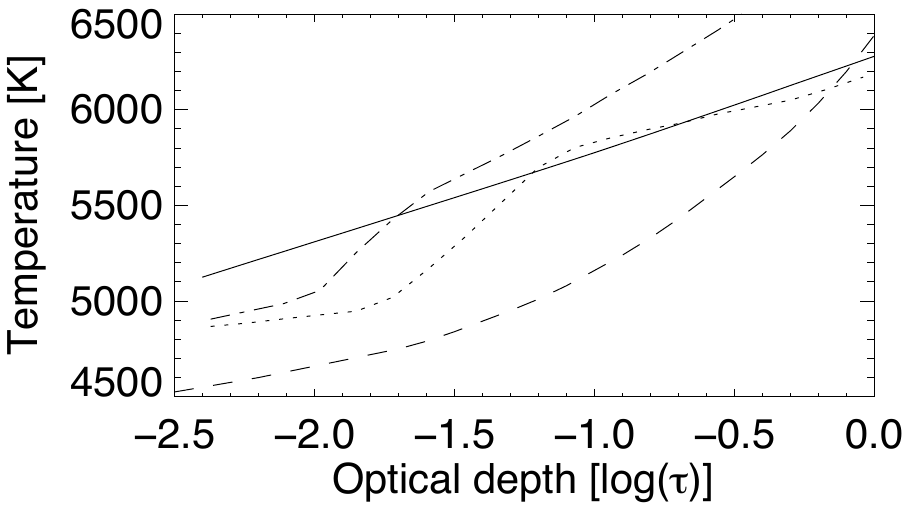}
   \end{minipage} 
   \caption{$\emph{Left}$: Distribution of temperature ratios of the $\log(\tau)=0$ and $-2.3$ layers of all $\emph{core}$ pixels. The $\emph{dotted}$ line shows the same ratio obtained from the plage flux tube model of \citet{solanki1992pla}. $\emph{Right}$: The $\emph{solid}$ line represents the temperature stratification of a typical $\emph{core}$ pixel obtained from the inversion. The $\emph{dotted}$ line follows the temperature stratification of an ideal plage flux tube model and the $\emph{dot-dashed}$ line indicates the temperature stratification of the network flux tube model. The $\emph{dashed}$ line depicts the temperature stratification of the HSRA.}
   \label{expanT}
   \end{figure*}

The temperatures at each of the three $\log(\tau)$ nodes is displayed in Fig. \ref{imgT} for the same FOV as in Fig. \ref{imgVel}. Fig. \ref{imgT}a corresponds to the temperature at $\log(\tau)=0$ and exhibits the familiar granulation pattern. The positions of $\emph{core}$ pixels are revealed by the black contour line in the images and show that they (the $\emph{core}$ pixels) are predominantly found within the comparatively cool intergranular lanes. Figures \ref{imgT}b $\&$ \ref{imgT}c display the temperature at $\log(\tau)=-0.9$ and $-2.3$, respectively. Both images indicate the high temperatures within MFCs when compared to the quiet Sun at these heights. Furthermore, the temperature map at $\log(\tau)=-2.3$ not only reveals the comparatively hot MFCs but also a reversed granulation pattern.\\
The higher temperatures within MFCs at $\log(\tau)=-0.9$ and $-2.3$, compared to the quiet Sun, are illustrated more quantitatively by the histograms in Fig. \ref{T}. At both those layers the average temperature is around 300 K higher within $\emph{core}$ pixels, with average temperatures of 5690 K and 5070 K at $\log(\tau)=-0.9$ and $-2.3$, respectively, than in quiet Sun pixels, where the average temperatures at the same $\log(\tau)$ heights are 5290 K and 4780 K, respectively. Fig. \ref{T}c further demonstrates that the average temperature in $\emph{canopy}$ pixels at $\log(\tau)=-2.3$ is only slightly lower than the temperatures of $\emph{core}$ pixels at the same height, with a mean of 5000 K. Only at $\log(\tau)=0$ is the average quiet Sun temperature higher, at 6410 K, than in the $\emph{core}$ pixels, which have a mean temperature of 6270 K. Quiet Sun pixels harbouring downflows ($\emph{dotted red}$ histogram in Fig. \ref{T}a) have a slightly lower mean temperature at 6240 K than MFCs, which are also located predominantly in downflowing regions. The MFC temperature histograms at $\log(\tau)=0$ have been artificially curtailed by the 5800 K threshold imposed at the beginning of the investigation to remove pores. However, the bulk of plage pixels is well above this threshold. Quiet Sun pixels in upflowing regions at $\log(\tau)=0$ ($\emph{dotted blue}$ histogram in Fig. \ref{T}a) have a mean temperature of 6590 K, well above the values of MFCs.\\  
The temperature gradient within $\emph{core}$ pixels was studied further, first, by taking the ratio of the $\log(\tau)=0$ and $-2.3$ temperatures. A histogram of these ratios is seen in the left panel of Fig. \ref{expanT}. The average ratio for $\emph{core}$ pixels is $0.81\pm0.02$, which demonstrates that the majority of $\emph{core}$ pixels have a very similar temperature stratification. The ratios were then compared to the temperature ratio obtained from a plage flux tube model derived by \citet{solanki1992pla}, which is shown by the $\emph{dotted}$ line in the left panel in Fig. \ref{expanT}. The temperature ratio of the model, which has a ratio of 0.79, agrees reasonably well with the inversion results. The thin flux-tube network model \citep{solanki1986}, has a ratio of 0.7 between the same $\log(\tau)$ heights and lies outside the histogram. The temperature stratification of MFCs studied in this investigation significantly deviate from the network model's prediction, which is not surprising given that we are studying strong plage.\\ 
The temperature stratification of various models is depicted in the right panel of Fig. \ref{expanT}. The $\emph{dotted}$ and $\emph{dot-dashed}$ curves represent the plage and network flux-tube models \citep{solanki1986,solanki1992pla}, while the $\emph{solid}$ line in the same panel shows the typical temperature stratification obtained from $\emph{core}$ pixels by the inversion. Whilst quantitatively the model and inversion results agree quite well, in particular at the three $\log(\tau)$ nodes, qualitatively there is an important difference. The temperature stratification of the model has a notable bend between $\log(\tau)=-1$ and $-1.5$, which is entirely absent from the inversion result. However, such a bend can only be reproduced by the inversion by employing at least four nodes, which would in turn introduce too many free parameters, when inverting only 2 spectral lines.\\   
Whilst on average the temperature stratification of a MFC follows that a thin flux tube model, a closer inspection of the MFC's cross-sections reveals that the temperature within a MFC is not uniformly distributed at all. The highest temperatures appear to be preferentially located at the edges of the MFCs. At $\log(\tau)=0$, Fig. \ref{imgT}a, the temperature gradients across a MFC are strongest, whilst higher layers display an ever more uniform distribution of temperatures. Nevertheless, even at $\log(\tau)=-2.3$, Fig. \ref{imgT}c, some areas within the MFCs have an enhanced temperature when compared to their immediate surroundings. These enhanced temperature areas can usually be traced through all three layers and become smaller in size in deeper layers, e.g. at $-242.5X$ and $-235Y$ or at $-238.5X$ and $-240Y$. The $\emph{white}$ areas in Fig. \ref{imgT}a indicate the pixels containing supersonic velocities (see Sect. 4.4). Many examples can be found in Fig. \ref{imgT} of a match between the location of supersonic velocities and a nearby (1-2 pixels away, but always within the MFC) local temperature enhancement. However, there is no one-to-one relationship between the two. At numerous locations throughout the plage region, e.g. at $-238X$ and $-237.5Y$ in Fig. \ref{imgT}, there is a clear local enhanced of the temperature across all three $log(\tau)$ layers, but no nearby supersonic velocity. Furthermore, no linear relationship seems to exist between the magnitude of the temperature enhancement (in any layer) and the magnitude of the nearby supersonic downflow. \\
   
\subsection{Inclination $\&$ Azimuth} 
\begin{figure*}
   \centering
   \includegraphics{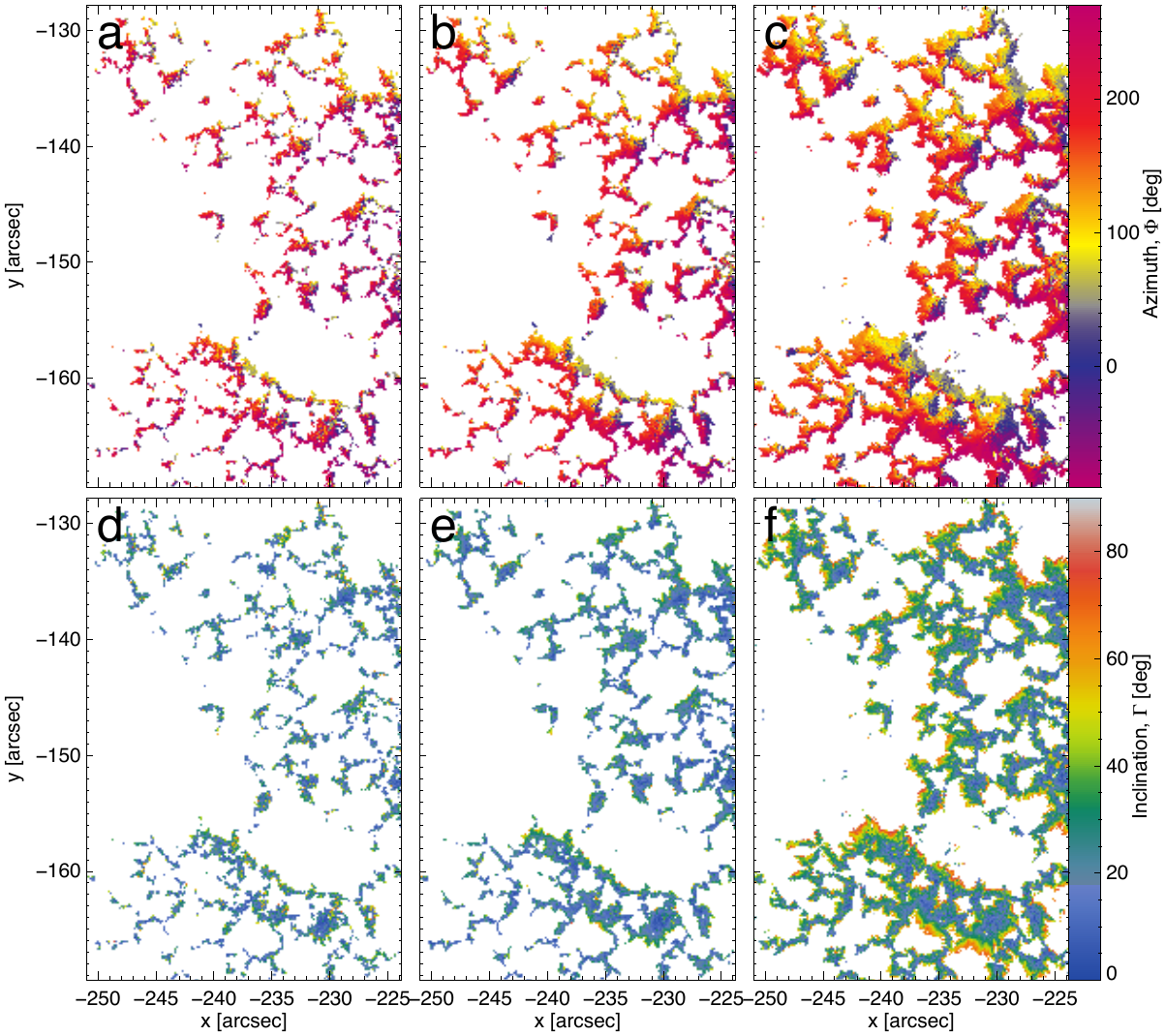}
   \caption{$\emph{a-c}$: Ambiguity resolved azimuths in local solar coordinates, $\Phi$, at $\log(\tau)=0, -0.9$ and $-2.3$ from left to right. North is up and corresponds to an angle of $90^{\circ}$. West is to the right and corresponds to an angle of $0^{\circ}$. $\emph{d-f}$: The local solar coordinate corrected inclinations, $\Gamma$, of the magnetic field after the azimuth ambiguity resolution at $\log(\tau)=0, -0.9$ and $-2.3$ from left to right. The inclinations and azimuths of pixels with $B$<300 G at $\log(\tau)=-2.3$ and $B$<700 G at $\log(\tau)=0$ $\&$ $-0.9$ are shown in white.}
    \label{genAziIncl}
    \end{figure*}

The inclinations of the magnetic field plotted in Figs. \ref{imgBIncl} are inclinations in the observer's frame of reference. Due to the small heliocentric angle ($\langle\theta\rangle=13^o$) a qualitative picture of the inclinations of plage magnetic features can still be gained from those figures. However, a conversion of these inclinations to local solar coordinates is necessary to find the inclination of the magnetic fields with respect to the solar surface.\\
The conversion of the inclinations and azimuths retrieved by the inversion to local solar coordinates is not straightforward. Whilst the inclination of the magnetic field is uniquely defined the azimuth has an inherent $180^\circ$ ambiguity \citep{unno1956}. Therefore, when converting to local solar coordinates one is forced to choose between one of two possible solutions for the magnetic field vector. Many codes, requiring various amounts of manual input, have been developed to solve the $180^\circ$ ambiguity. The reader is directed to  \citet{metcalf2006} and \citet{leka2009} for overviews. An additional challenge facing these codes is that they generally use the output of an ME inversion as an input. The output of the ME inversion does not contain any information on the change of the magnetic field vector over the formation height of the inverted absorption line. The SPINOR code, however, provides this information, which in turn allows the canopies of magnetic features to be identified as is shown in  Fig. \ref{imgcanopy}. The azimuths retrieved by the inversion were spatially very smooth in all the nodes, indicating that the azimuth was well defined in the regions containing magnetic fields.\\ 
The $\emph{canopy}$ pixels shown in Fig. \ref{imgcanopy} form continuous rings around the various $\emph{cores}$. By assuming that the magnetic field in each $\emph{canopy}$ pixel originates from the largest patch of $\emph{core}$ pixels in its immediate vicinity, the direction of the magnetic field vector of each $\emph{canopy}$ pixel can be determined unambiguously as long as the polarity of the corresponding $\emph{core}$ patch is known. The polarity of a patch of $\emph{core}$ pixels can be determined easily from their Stokes $V$ spectra. In essence, the magnetic field vector of a $\emph{canopy}$ pixel points towards a patch of $\emph{core}$ pixels if it has a negative polarity and, in turn, away from it if the patch of $\emph{core}$ pixels has a positive polarity. With the help of this process we were able to resolve the azimuth ambiguity of the canopy pixels without having to make any further assumptions on the properties of the magnetic field. This process was repeated for each $\emph{canopy}$ pixel individually. Since the canopies of the various plage features are greatest at $\log(\tau)=-2.3$, the magnetic field vector of the $\emph{canopy}$ pixels was determined at this $\log(\tau)$ height.\\
Once the magnetic field vectors of the $\emph{canopy}$ pixels were determined, the vectors of the $\emph{core}$ could be obtained using an acute-angle method. This method works by performing dot products, using the two possible vectors in an undetermined pixel, with those surrounding pixels whose vector was already determined. The dot products corresponding to each of the two possible solutions were then summed. The vector associated with the smallest sum was subsequently selected as the correct vector for that pixel. The pixels that were surrounded by the largest number of already determined field vectors were given preference.\\
Once the magnetic field vectors of both the $\emph{canopy}$ and $\emph{core}$ pixels at $\log(\tau)=-2.3$ were known, the vectors at $\log(\tau)=0$ $\&$ $-0.9$ could also be determined. The now known vector at $\log(\tau)=-2.3$ in each pixels was used to perform dot products with the two possible vectors in the next lower $\log(\tau)$ layer. The vector with the smaller dot product was subsequently chosen as the correct magnetic field vector. The $180^{\circ}$ ambiguity of the magnetic field vector at $\log(\tau)=0$ $\&$ $-0.9$ was removed for only those pixels where $B$>700 G in either layer. The resolution of all the vectors is entirely automatic and only the definition of the $\emph{core}$ and $\emph{canopy}$ pixels for the initial input is manual, but followed the definition given in Sect. 3. Also, no smoothing of the azimuths is performed at any point. The converted inclinations and azimuths in local solar coordinates after the resolution of the $180^{\circ}$ ambiguity are plotted in Fig. \ref{genAziIncl}.\\
Figures \ref{genAziIncl}a-c show the resolved azimuths, $\Phi$, at all three nodes. Several 'azimuth centres' \citep{martinez1997} can be readily identified, in particular at $\log(\tau)=-2.3$. In combination with Figures \ref{genAziIncl}d-f it can be seen that most of the 'azimuth centres' have vertical fields in their $\emph{cores}$ that become more horizontal towards the edges of a MFC. 'Azimuth centres' tend to be either relatively isolated features or large ones. Most of the MFCs tend to be elongated with the field expanding roughly perpendicular away from the long axis of the structure over most of its length and directed radially away at the ends.\\
     \begin{figure}
   \centering
   \includegraphics[width=7cm]{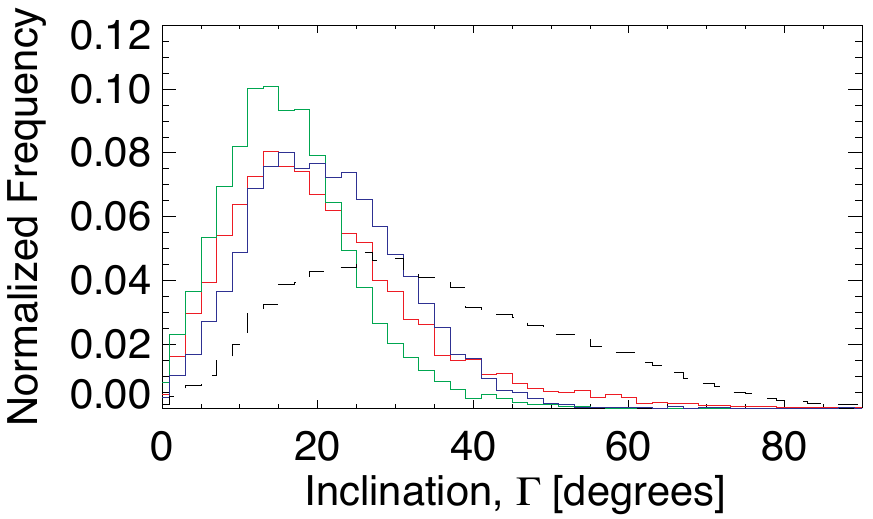}
      \caption{Histograms of magnetic field inclination of MFCs relative to the solar surface normal, $\Gamma$. The three coloured histograms were obtained using $\emph{core}$ pixels, where \emph{red} refers to $\log(\tau)=0$, \emph{green} shows to $\log(\tau)=-0.9$ and \emph{blue} refers to $\log(\tau)=-2.3$. The $\emph{dashed}$ histogram depicts $\Gamma$ of $\emph{canopy}$ pixels at $\log(\tau)=-2.3$.}
         \label{genIncl}
   \end{figure}
A more quantitive picture of the general inclinations of the $\emph{core}$ pixels can be obtained through their histograms, depicted in Fig. \ref{genIncl}. The distributions of the inclinations have their peak between $10^\circ$ and $15^\circ$ for all $\log(\tau)$ nodes. The mean inclination for each $\log(\tau)$ layer is $22^{\circ}$, $18^{\circ}$ and $23^{\circ}$ with decreasing optical depth. The median values have a similar progression with optical depth, taking values of $19^{\circ}$, $17^{\circ}$ and $21^{\circ}$ respectively. Fig. \ref{genIncl} also reveals that the $\emph{canopy}$ pixels are significantly more horizontal, with a peak at $25^\circ$ and a very extended tail reaching up to $90^{\circ}$. The mean inclination for the $\emph{canopy}$ fields is $39^{\circ}$ with a median of $36^{\circ}$, which demonstrates quantitatively the more horizontal nature of the $\emph{canopy}$ when compared to the $\emph{core}$ fields. The largest inclinations are found at the edges of the canopies as expected for an expanding flux tube or flux sheet.\\
   \begin{figure}
   \centering
   \includegraphics[width=7cm]{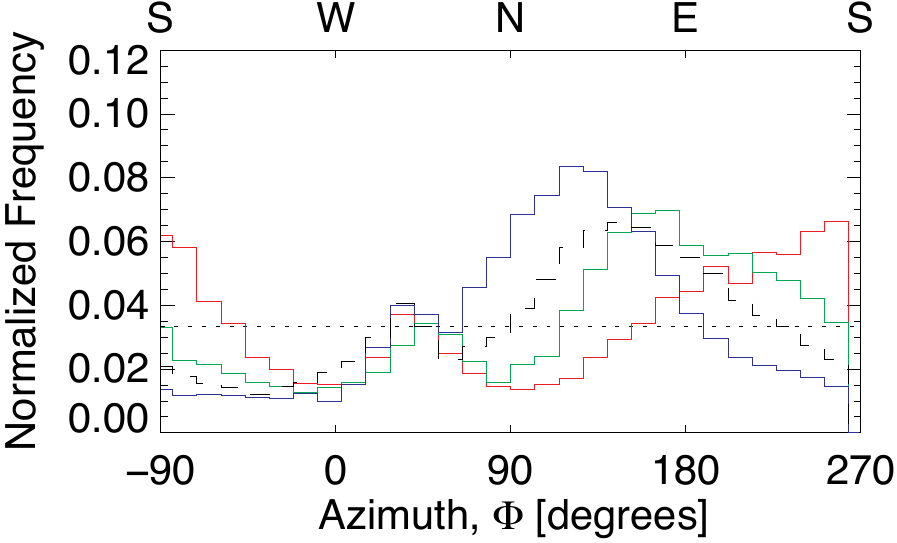}
      \caption{Histograms of $\Phi$ found in plages. The three coloured histograms were obtained using $\emph{core}$ pixels, where \emph{red} refers to $\log(\tau)=0$, \emph{green} shows to $\log(\tau)=-0.9$ and \emph{blue} refers to $\log(\tau)=-2.3$. The $\emph{dashed}$ histogram shows $\Phi$ of $\emph{canopy}$ pixels at $\log(\tau)=-2.3$. The $\emph{dotted}$ line represents a homogeneous distribution.}
         \label{genAzi}
   \end{figure}
The histograms of $\Phi$ of $\emph{core}$ pixels are depicted in Fig. \ref{genAzi}. None of the four distributions are homogeneous and show a consistent under-representation of the $W$ and partly the $N$ directions. These two directions are, however, expected to be under-represented due to the viewing geometry, as the region was located in the $SE$ at the time of the observation. The azimuth distributions from MFCs found at the northern edge of the field of view show a more homogeneous distribution, as expected.\\
Fig. \ref{genAzi} also shows that the peak of each azimuth distribution is shifted with respected to other distributions, suggesting that the direction of the magnetic field vector of individual pixels in MFCs appears to rotate with height. Several tests were carried to determine the nature of this rotation, after which a solar origin as well as an inversion based error seem unlikely. Several instrumental effects such as cross-talks or differences in the spectral dispersion between the Stokes parameters were found to be capable in causing the observed rotation. However, a more in depth investigation regarding this matter is required. \\ 

\subsection{Effect of the sunspot}
\begin{figure*}
   \centering
   \begin{minipage}{0.5\linewidth}
   \centering
   \includegraphics[width=7cm]{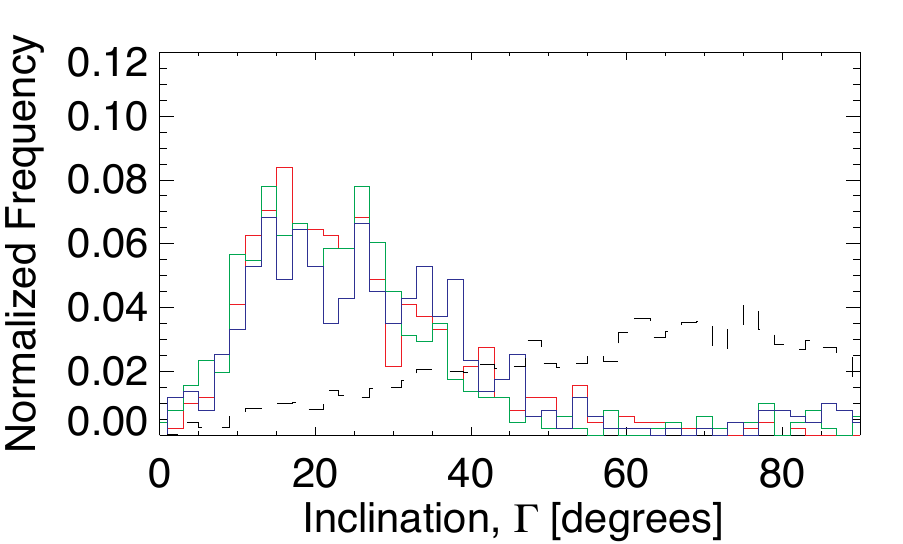}
   \end{minipage}%
   \begin{minipage}{0.5\linewidth}
   \centering
   \includegraphics[width=7cm]{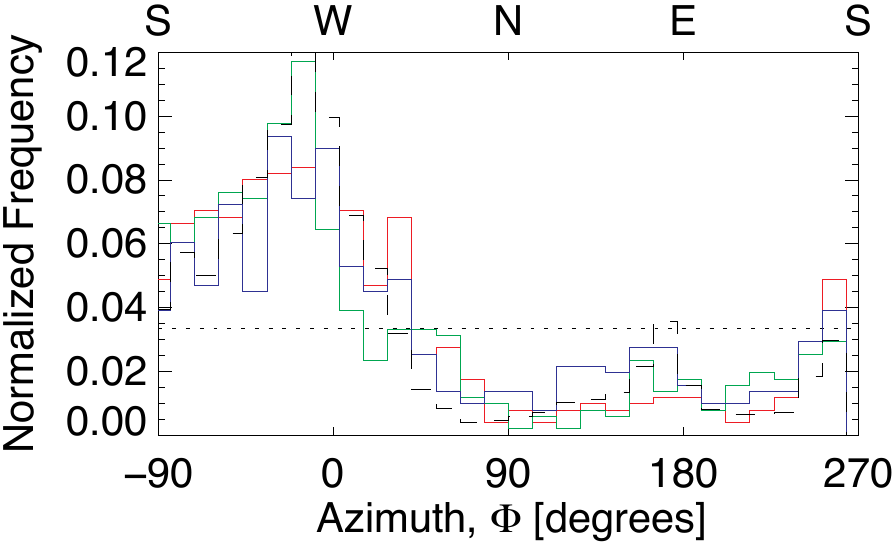}
   \end{minipage} 
   \caption{$\emph{Left}$: Histograms of $\Gamma$ found in plage around $-220X$, $-170Y$. The three coloured histograms were obtained using $\emph{core}$ pixels, where \emph{red} refers to $\log(\tau)=0$, \emph{green} shows to $\log(\tau)=-0.9$ and \emph{blue} refers to $\log(\tau)=-2.3$. The $\emph{dashed}$ histogram represents $\theta$ of $\emph{canopy}$ pixels at $\log(\tau)=-2.3$.
   $\emph{Right}$: Histograms of $\Phi$ found in plages around $-220X$, $-170Y$. The three colours and the $\emph{dashed}$ line have the same significance as in the graph on the left in this figure. The $\emph{dotted}$ line represents a homogeneous distribution.}
   \label{nrspot}
   \end{figure*}

The majority of the MFCs show no obvious and conclusive influence of the nearby sunspot's magnetic field, which stretches well beyond the spot's visible boundary in the form of a low-lying magnetic canopy, as can be seen in Fig. \ref{imgcanopy}. Similar extended sunspot canopies were already found in earlier studies \citep[e.g.][]{giovanelli1982,solanki1992expan,solanki1994}. A few MFCs in the FOV, however, are noticeably influenced by the sunspot. The most striking of these features is located at $-217X$, $-125Y$  in Figs. \ref{imgBIncl} \& \ref{imgcanopy}, where an extensive loop system can be identified. These loops have horizontal fields and connect several pores with positive polarity to the negative polarity sunspot. The magnetic field strengths found in this loop system can reach values as high as 1000 G at $\log(\tau)=-2.3$ in a few places and Fig. \ref{imgcanopy} indicates that they are suspended above relatively field-free gas, since they are identified as $\emph{canopy}$ pixels.\\
MFCs close to the sunspot also posses highly deformed canopies, which are elongated towards the spot if the MFC has the opposite polarity of the spot. Such MFCs can be seen at $-220X$,$-170Y$ and histograms of their inclinations and azimuths are presented in Fig. \ref{nrspot}.
The field in the $\emph{core}$ pixels of these MFCs is more inclined than on average; compare with Fig. \ref{genIncl}. The mean inclinations of the field at the three layers from $\log(\tau)=0$ to $\log(\tau)=-2.3$ are $31^{\circ}$, $32^{\circ}$ and $35^{\circ}$, respectively. These average inclinations are about $10^{\circ}$ larger than for MFCs found further away from the spot. In particular the inclinations of the $\emph{canopy}$ pixels in Fig. \ref{nrspot} reveal the effect of the sunspot upon these magnetic fields even more strongly. The mean inclination of the canopy fields is $56^{\circ}$ and the median value is $57^{\circ}$, which is more than $15^{\circ}$ larger than the average in Fig. \ref{genIncl}. The azimuth distributions in Fig. \ref{nrspot} clearly display the influence of the sunspot as all the distributions both from the $\emph{core}$ and $\emph{canopy}$ pixels show a clear preferred orientation towards the spot. It is likely that the canopies of these MFCs interact with the sunspot's canopy \citep{solanki1996}, even if the fields of both magnetic structures do not appear to be directly connected over the $\log(\tau)$ range considered in this investigation.\\
From the examples given in this section the following picture emerges. Those magnetic features that are in the immediate vicinity of the sunspot, i.e. if there are no further kG magnetic fields in between them and the spot, show a clear deformation of their canopy and have inclination and azimuth distributions that either predominantly point towards or away from the spot, depending on the polarity. The spot's influence on the orientation of magnetic features could be observed up to $20"$ away from the sunspot's outer penumbral boundary, provided there were no other magnetic features in between. Any magnetic feature situated behind this 'first row' of kG features is effectively shielded from the spot in the photosphere and then behaves closer to an isolated magnetic feature, affected only by its nearest neighbours.\\

\subsection{Weak opposite polarity fields next to MFCs}
   \begin{figure*}
   \centering
   \includegraphics[width=18cm]{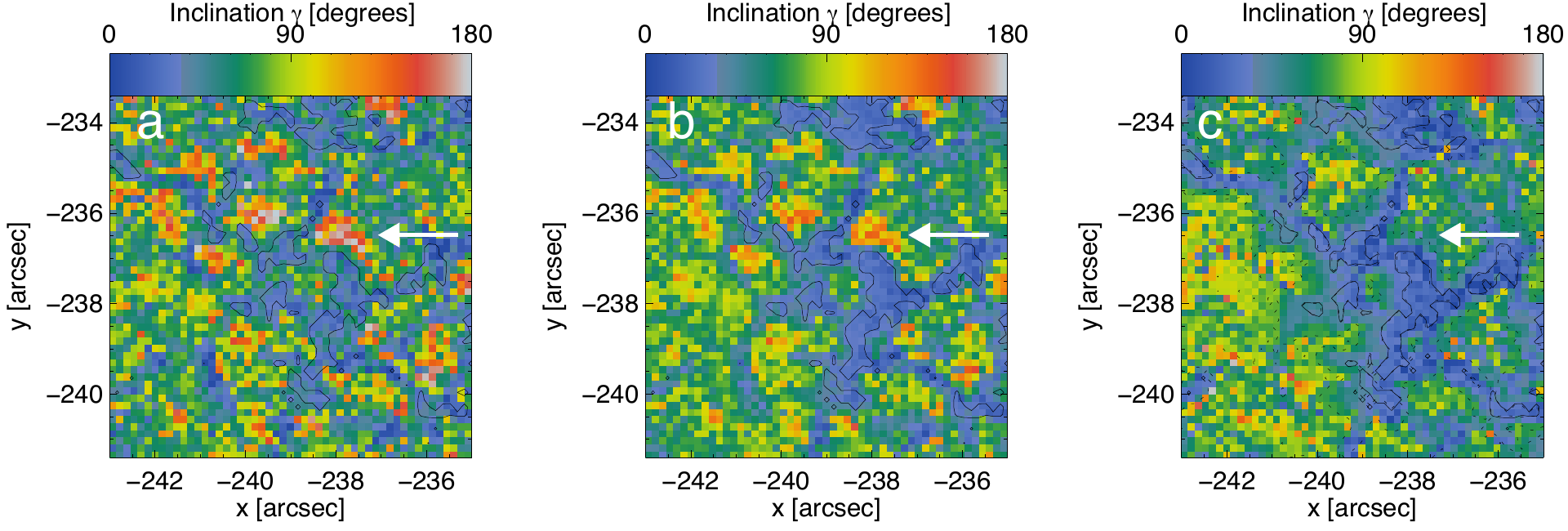}
   \caption{$\emph{a-c}$: LOS inclination, $\gamma$, at $\log(\tau)=0, -0.9$ and $-2.3$, respectively. The black contour lines encompass $\emph{core}$ pixels. The arrows point to the location of the weak opposite polarity at each height.}
    \label{imgIncl}
    \end{figure*}
\begin{figure}
   \centering
   \includegraphics[width=9cm]{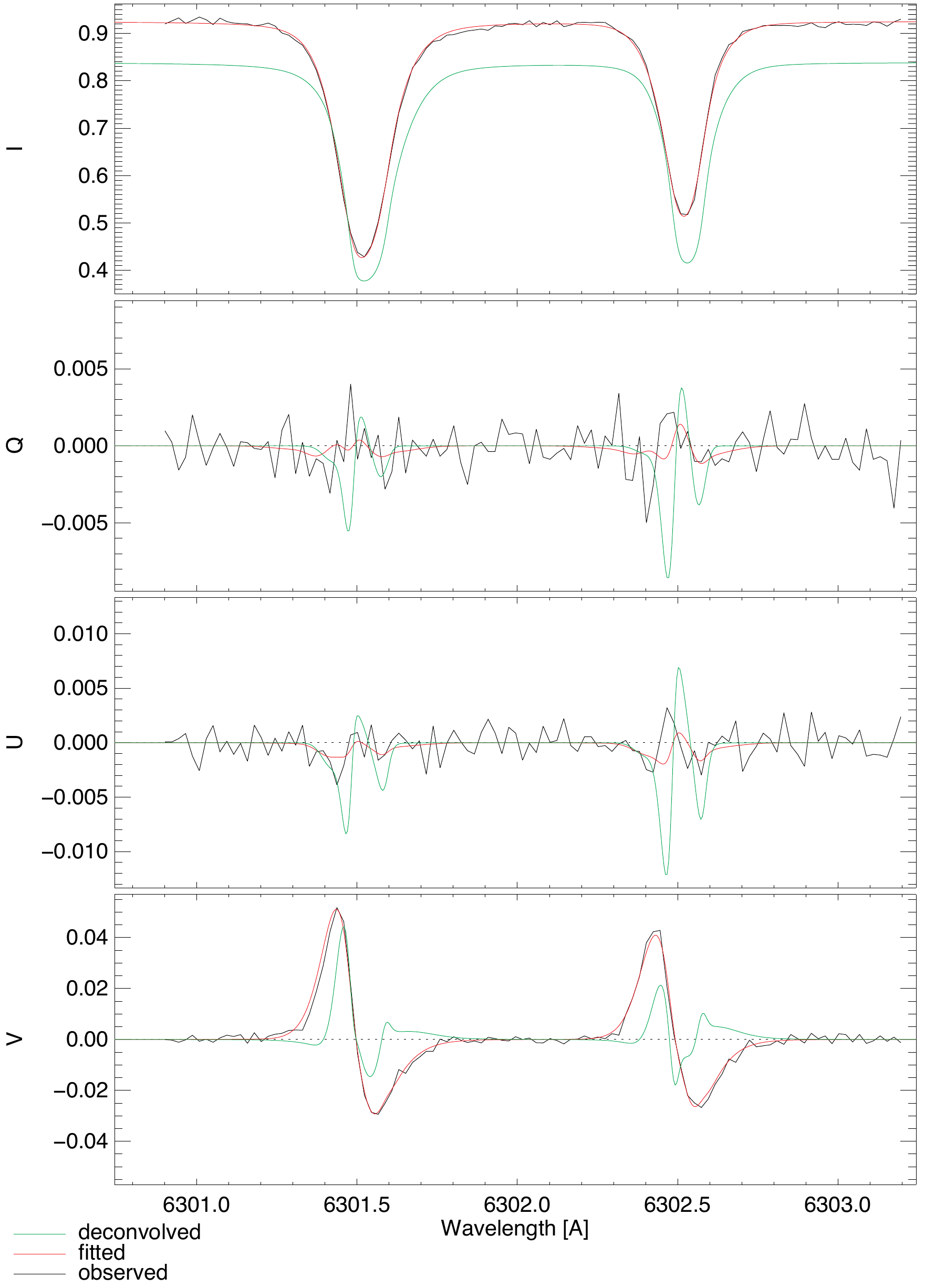}
      \caption{Stokes spectra of a typical $\emph{canopy}$ pixel harbouring a small opposite polarity magnetic field at $\log(\tau)=0$ and $-0.9$. The $\emph{black}$ spectra correspond to the original SOT/SP observation, the $\emph{red}$ curves display the spectra fitted by the inversion, and the resultant spectra after the spatial deconvolution are coloured $\emph{green}$.}
         \label{spectra}
   \end{figure}

Figures \ref{imgIncl}a-c show the LOS inclinations of the magnetic field over a small FOV (same as Figs. \ref{imgVel}, \ref{imgT} $\&$ \ref{imgmicro}). In this blow-up the fields in the MFCs ($\emph{black}$ contour lines) display mainly vertical orientations, shown in $\emph{blue}$, but a closer inspection of the images corresponding to the $\log(\tau)=0$ $\&$ $-0.9$ layers exhibit that many MFCs are adjoint by magnetic patches with an inclination opposite to the MFC, as can be seen from their $\emph{red}$ colour, e.g. at $-237.5X$ and $-238.5Y$. Further examination of Fig. \ref{imgIncl} reveals that these small opposite polarity patches are hidden beneath the canopy of the nearest MFC. As a result the small $\emph{red}$ patches are absent in Fig. \ref{imgIncl}c. The vast majority of these small opposite polarity patches have $B$<100 G at $\log(\tau)=0$ $\& -0.9$, whereas the canopies above them have $B$>300 G. This means that the Stokes spectra of these pixels are dominated by the canopy fields. Only the deconvolved Stokes profiles returned by the inversion, as displayed in Fig. \ref{spectra}, show a small polarity reversal in the wings of the Stokes $V$ profile.\\
   \begin{figure*}
   \centering
   \begin{minipage}{0.5\linewidth}
   \centering
   \includegraphics[width=7cm]{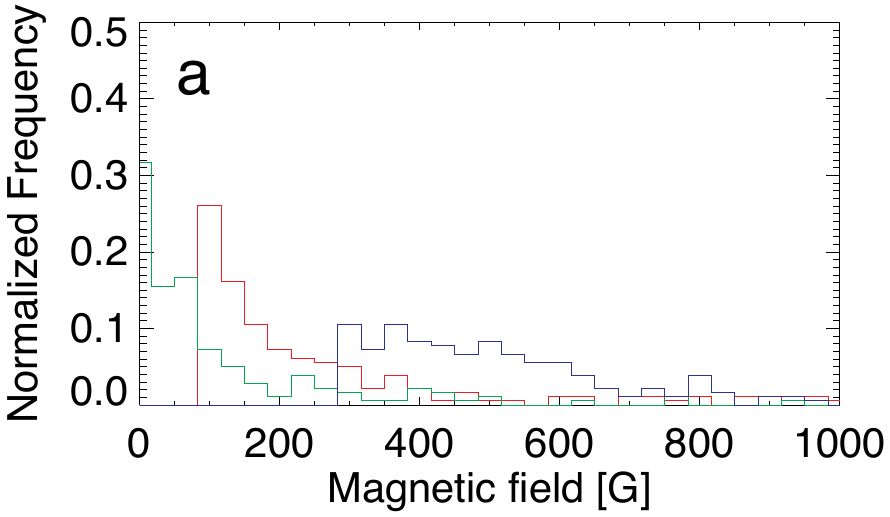}
   \end{minipage}%
   \begin{minipage}{0.5\linewidth}
   \centering
   \includegraphics[width=7cm]{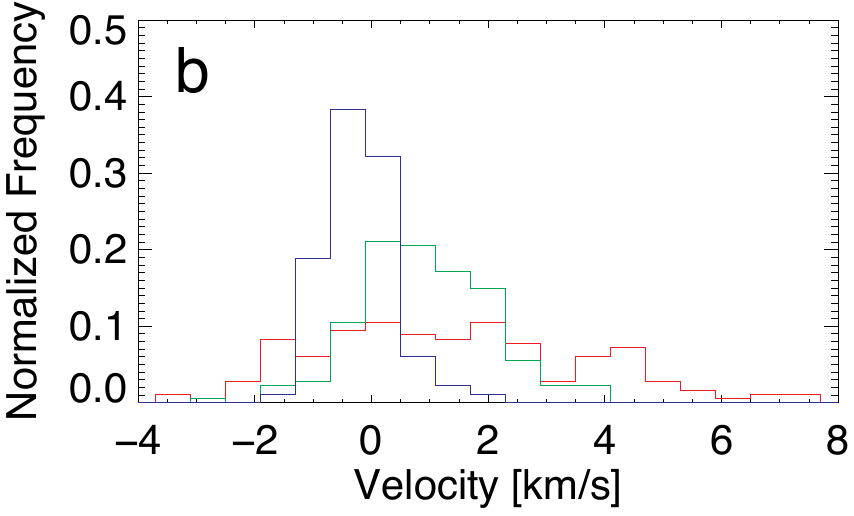}
   \end{minipage}   
   \caption{ $\emph{a}$: Histograms of the magnetic field strengths in pixels featuring weak opposite polarities. The colours $\emph{red}$, $\emph{green}$, $\emph{blue}$ refer to $\log(\tau)=0, -0.9$ $\& -2.3$, respectively. $\emph{b}$: Histograms of the LOS velocities in the same pixels. The colour code is identical.}
   \label{reversal}
   \end{figure*}     
Nonetheless a number of pixels in these small opposite polarity patches have field strengths in excess of 100 G at $\log(\tau)=0$. These pixels were analysed further by producing histograms of their field strengths and LOS velocities, which are depicted in Fig. \ref{reversal}a $\&$ b, respectively. At $\log(\tau)=0$ the mean and median of the histogram are only 280 G and 190 G, respectively. These values drop to 160 G and 70 G at $\log(\tau)=-0.9$ before rising to the considerably higher strengths of the $\emph{canopy}$ fields at $\log(\tau)=-2.3$. The histograms in Fig. \ref{reversal}b, corresponding to the lower two $\log(\tau)$ layers, demonstrate that these weak opposite polarity fields are predominantly located in the strong downflows surrounding the MFCs. The mean and median velocities of these pixels at $\log(\tau)=0$ are 1.6 km/s and 1.3 km/s, respectively. At $\log(\tau)=-0.9$ the mean and median velocities are 1.3 km/s and 1.2 km/s. The velocities at $\log(\tau)=-2.3$ are, as expected, the same on average to those seen in the $\emph{dashed}$ histogram in Fig. \ref{genVel}, since at this height the LOS velocity inside the MFC is sampled and, hence, mostly weak flows are seen. The flux stored in these opposite polarity patches is tyically below $1\%$ of the flux of their parent MFCs.

\subsection{Microturbulence}
The inclusion of micro- and macro-turbulent velocities is common when fitting photospheric absorption lines and is an indication of unresolved fine structure \citep[e.g.][]{lites1973,holweger1978,solanki1986}. Quiet Sun Hinode SOT/SP observations containing only weak magnetic fields have been fitted without the inclusion of a microturbulence term \citep{socas2011}. In areas with kG magnetic fields, however, an excess of turbulent velocities was found by \citet{solanki1986}. The total rms value for turbulent velocities has a range of 1.0 km/s to 3.5 km/s in areas containing kG magnetic fields, depending on the spectral line and on the line strength. \citet{zayer1989} found, using lines in the infrared, that rms turbulent velocities between 3.0 km/s and 3.5 km/s were necessary to fit the observed profile shapes. The dependence on spectral line strength suggests a height dependent turbulent velocity. Therefore, we carried out the inversions with a height dependent microturbulence, while forcing the macroturbulence to zero. This approach turned out to give satisfactory fits to the line profiles. Restricting ourselves to microturbulence alone is also in line with the improved spatial resolution and stable observing conditions of the Hinode satellite. Any remaining non-thermal broadening in the spectral line profiles is assumed to be caused by unresolved velocities within the resolution element. A mixture of up- and downflows with a large correlation length along the LOS is less likely at high resolution, than in the low resolution data analysed in the earlier investigations.\\
    \begin{figure*}
   \centering
   \includegraphics[width=14cm]{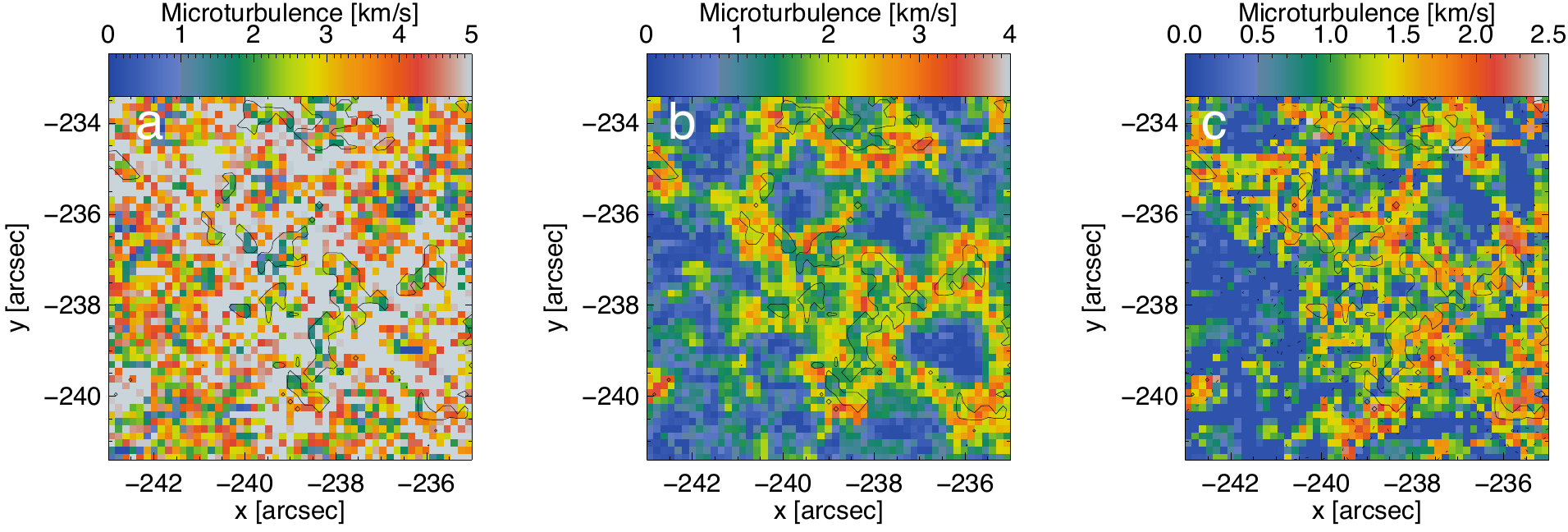}
   \caption{$\emph{a-c}$: Microturbulent velocities retrieved by the inversion at $\log(\tau)=0, -0.9$ and $-2.3$ from left to right. Note the different colour scale for each frame. The black contour encompasses $\emph{core}$ pixels in all images. The $\emph{dotted}$ lines (in panel c) display $\emph{canopy}$ pixels.}
    \label{imgmicro}
    \end{figure*} 
Figure \ref{imgmicro} displays the microturbulent velocities, $\xi_{mic}$, retrieved by the inversion at the three $\log(\tau)$ heights. In all three panels areas with an increased $\xi_{mic}$ are found to coincide with areas of strong magnetic fields. The quietest areas are found to have the lowest turbulent velocities, supporting the finding of \citet{socas2011} that turbulent velocities are low in the quiet Sun. In particular, Fig. \ref{imgmicro}b reveals that although $\emph{core}$ pixels display enhanced microturbulence, the largest turbulent velocities are preferentially situated at the edges of a magnetic feature, forming a narrow halo or ring around MFCs. A closer inspection of Fig. \ref{imgmicro}c reveals that the same is also true at $\log(\tau)=-2.3$, although the halo is less well marked at this height. At $\log(\tau)=0$ the halo of large microturbulence around the MFCs is more extended, although large microturbulent velocities are found everywhere. In this inversion the upper limit for the microturbulence was set at 5 km/s. If a higher upper limit was set then microturbulent velocities of up to 9 km/s were retrieved in the halo by the inversion at $\log(\tau)=0$.\\ 
   \begin{figure*}
   \centering
   \begin{minipage}{0.5\linewidth}
   \centering
   \includegraphics[width=7cm]{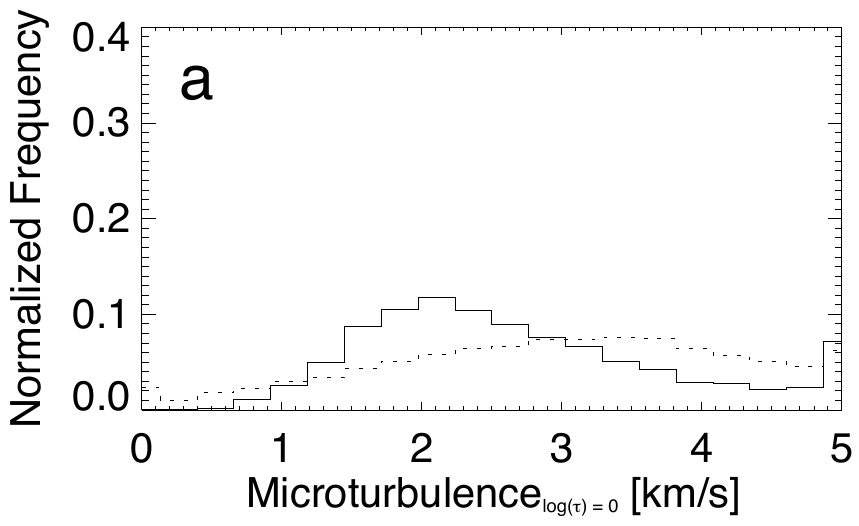}
   \end{minipage}%
   \begin{minipage}{0.5\linewidth}
   \centering
   \includegraphics[width=7cm]{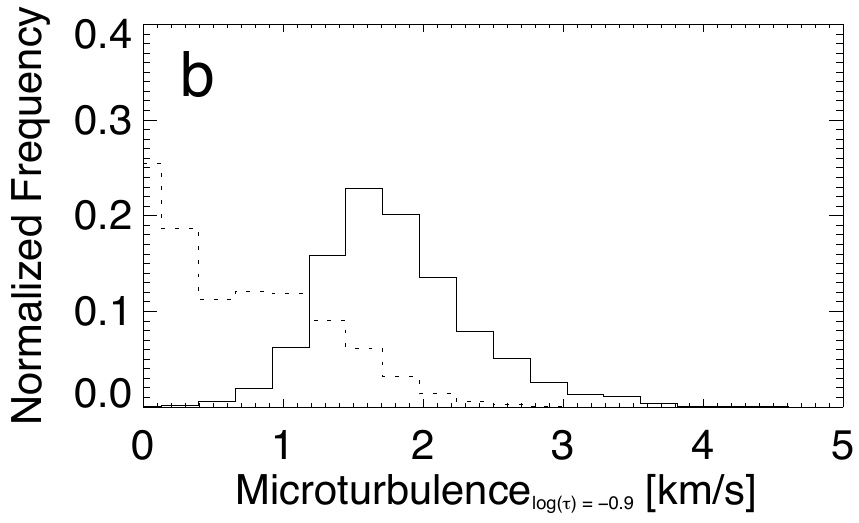}
   \end{minipage}
   \begin{minipage}{0.5\linewidth}
   \centering
   \includegraphics[width=7cm]{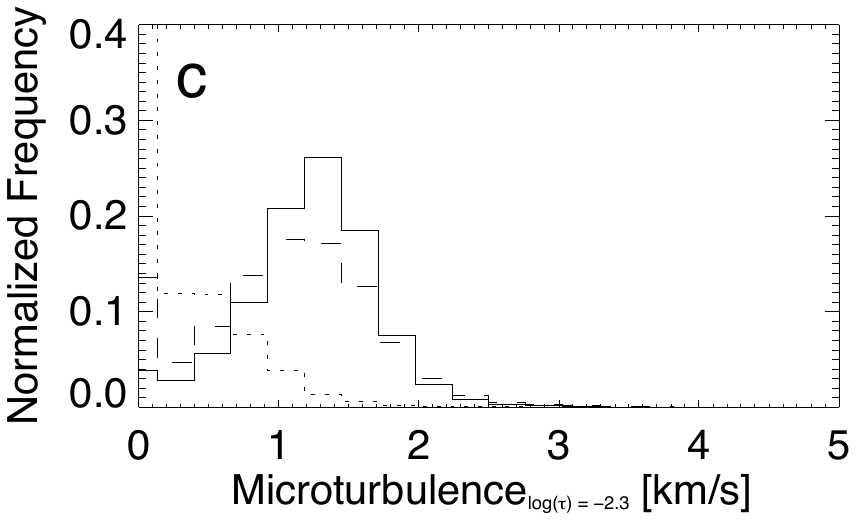}
   \end{minipage}    
   \caption{ Histograms of microturbulent velocities in $\emph{core}$ pixels, $\emph{solid}$ and the quiet Sun, $\emph{dotted}$, at $\log(\tau)=0$, $\emph{a}$, at $\log(\tau)=-0.9$, $\emph{b}$, and at $\log(\tau)=-2.3$, $\emph{c}$. The $\emph{dashed}$ histogram in panel, c at $\log(\tau)=-2.3$, represents the microturbulent velocities of $\emph{canopy}$ pixels.}
   \label{micro}
   \end{figure*}
   \begin{figure}
   \centering
   \includegraphics[width=7cm]{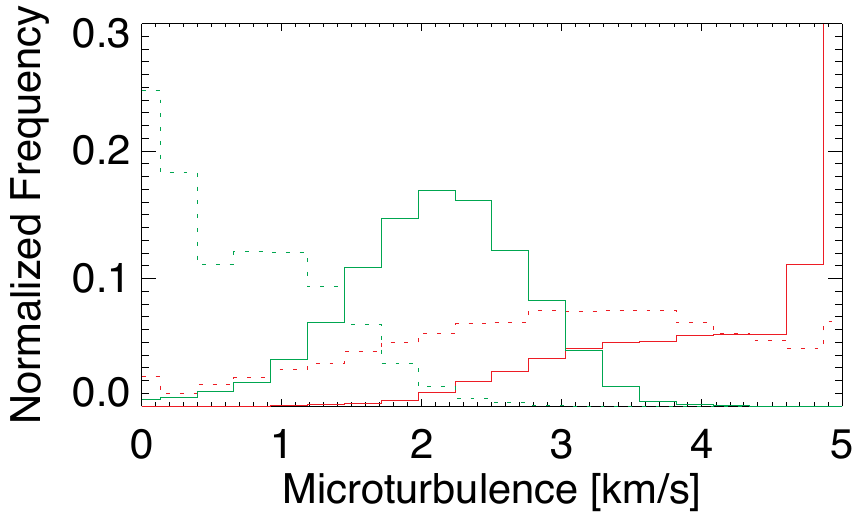}
      \caption{Histograms of microturbulent velocities in pixels immediately surrounding $\emph{core}$ pixels, same as Fig. \ref{Velring}, at $\log(\tau)=0$ in $\emph{red}$ and $\log(\tau)=-0.9$ in $\emph{green}$. The $\emph{dotted}$ histograms show the microturbulent velocities in the quiet Sun for the same layers.}
         \label{Microring}
   \end{figure}   
The distribution of $\xi_{mic}$, found in $\emph{core}$ pixels, are given by the $\emph{solid}$ histograms in Fig. \ref{micro}. These distributions can be compared to $\emph{dotted}$ histograms depicting $\xi_{mic}$ obtained from quiet Sun pixels. The differences in $\xi_{mic}$ between the quiet Sun and MFCs, suggested in Fig. \ref{imgmicro}, are confirmed by Fig. \ref{micro}. At $\log(\tau)=0$ $\xi_{mic}$ in the quiet Sun is on average slightly larger, with a mean $\xi_{mic}$ of 3.1 km/s, than in MFCs, which have an average velocity of 2.8 km/s. In higher layers the micro- turbulence in MFCs is significantly larger than in the quiet Sun. Within the MFCs the average $\xi_{mic}$ are 1.9 km/s and 1.3 km/s at $\log(\tau)=-0.9$ and $\log(\tau)=-2.3$, whilst in the quiet Sun the average $\xi_{mic}$ values are 0.8 km/s and 0.4 km/s in the two layers, respectively. Note that the canopy pixels require a similar microturbulence as the $\emph{core}$ pixels at the same optical depth. Fig. \ref{micro} also reveals that in the upper two layers there are many pixels in the quiet Sun which require no microturbulent broadening at all. \\
The microturbulent velocities found in pixels, which are immediately adjacent to $\emph{core}$ pixels are displayed in Fig. \ref{Microring}. The pixels used in this figure are identical to the ones used in Fig. \ref{Velring}, and are, therefore, the pixels featuring the highest LOS downflow velocities. Again, only the microturbulence in the lower two $log(\tau)$ layers are shown in Fig. \ref{Microring} (due to the overlying canopy at $\log(\tau)=-2.3$). The average and median microturbulent velocities are 4.4 km/s and 5.0 km/s at $\log(\tau)=0$ and 2.2 km/s and 2.3 km/s at $\log(\tau)=-0.9$. These values are higher than those in the quiet Sun and also than those found within the $\emph{core}$ pixels, demonstrating that the highest microturbulence is located at the edges of MFCs. Both the average and median values of $\xi_{mic}$ at $\log(\tau)=0$ in the surrounding, downflowing pixels are likely to be lower limits due to the upper bound of 5 km/s set on the microturbulence in the inversion.\\
The microturbulence retrieved by the inversion could, but does not necessarily imply the existence of turbulence or unresolved convective processes taking place within the magnetic features. It may point to unresolved waves in the MFC (e.g. surface waves could account for the higher $\xi_{mic}$ near the boundaries of the MFCs), or it may be due to unresolved horizontal velocity gradients strongest at the boundaries. Another possibility may be, at least in part, a signal of magnetic reconnection between the opposite polarity fields found in this study (see Sect. 4.9). Although it cannot be completely ruled out that inaccuracies in the damping constants of Fe I 6301.5 $\AA$ and 6302.5 $\AA$ \citep{anstee1995,barklem1997,barklem1998} may contribute to the deduced $\xi_{mic}$, such inaccuracies are unable to explain the excess in $\xi_{mic}$ in the magnetic features, since the spectral lines are significantly weakened there, so that damping becomes less important. 
   
\section{Discussion}

In the preceding sections we sought to ascertain some of the characteristic properties of the magnetic field in solar plage. Stokes $I$, $Q$, $U$ and $V$ profiles of AR 10953 observed by the SOT/SP aboard Hinode were analysed using the SPINOR inversion code \citep{frutiger2000}, extended by \citet{vannoort2012}, to perform a spatially coupled 2D inversion, which removes the influence of the instrument's PSF in parallel with inverting the data.\\
The inversion was able to retrieve and reproduce many of the previously deduced characteristic properties of magnetic flux concentrations (MFCs) in plage regions. The typical magnetic field strength of MFCs was found to be in the kG range and took on an average value of 1520 G at $\log(\tau)=-0.9$, where the 6302 $\AA$ line pair is most sensitive to the magnetic field. Similar photospheric magnetic field values were found previously by \citet[e.g.][]{wiehr1978,zayer1990,keller1990,rueedi1992,rabin1992,lin1995,martinez1997}. These values are, however, somewhat larger than those found using the Fe I 5250.2/5247.1 $\AA$ line pair \citep[e.g.][]{stenflo1985}, which is likely related to differences in the heights of formation of the two line pairs. Since no magnetic filling factor is introduced the obtained $B$ values are lower limits to the true field strengths. The magnetic field is found to drop more slowly with height when compared to a thin-tube (zeroth order) model. This lack of agreement stems from a lack of flux conservation with height, by as much as $20\%$, that primarily affects the upper $\log(\tau)$ node in the inversion. If this increase in magnetic flux is compensated then the stratification of the strongest magnetic fields with strengths higher than 2000 G, found at the centres of MFCs, do agree with the thin flux-tube model. When only the lower two $\log(\tau)$ layers are considered the model and the inversion also agree well. For magnetic fields $<2000$ G at $\log(\tau)=0$ other effects start to play a role.\\
The expansion of five isolated MFCs was compared to the ideal expansion of two $0^{th}$ order thin flux tube models. The expansion of the selected MFCs and the models agreed well, supporting previous results reported by \citet{pietarila2010expan}. At the highest inverted layers the field did expand somewhat less rapidly than in the model, however, probably due to the interactions (merging) with other magnetic features. The inversion demonstrates that the majority of MFCs in the strong plage investigated here merge with neighbouring MFCs already in the middle photosphere. Above the merging height $B$ drops more slowly or not at all and the field also does not expand very much \citep{pneuman1986,steiner1986}. The fact that a zeroth order flux tube is too simple to describe a MFC \citep{yelles2009} and expands more rapidly than more realistic models that include curvature forces may also contribute.\\
The LOS velocities obtained from the inversion show that the bulk of the magnetized gas in MFCs is essentially at rest and shows only weak downflows typically on the order of 200 m/s at $\log(\tau)=-0.9$, which agrees with the results of \citet{solanki1986,martinez1997}. Some of the $\emph{core}$ and $\emph{canopy}$ pixels show upflows or downflows of up to 1 km/s, which were also observed by \citet{langangen2007}. They are in most cases found within the MFCs and could be the result of oscillations or other transient events occurring within MFCs. The MFCs are each surrounded by a ring of strong downflows, which can be readily observed at the $\log(\tau)=0$ and $-0.9$ nodes in Figs. \ref{imgVel} $\&$ \ref{imgVelzoom}. These downflow rings were also seen to shift outwards in higher $\log(\tau)$ layers as the MFC expands. A downflow ring has an average velocity of 2.4 km/s at $\log(\tau)=0$, but parts of the ring achieve supersonic velocities up to 10 km/s, corresponding to a Mach number of 1.25, in the same $\log(\tau)$ layer. On average $2.5\%$ of a MFC's ring contains downflows reaching supersonic velocities. Some of these downflows appear to overlap with the magnetic field, which could be the signature of entrainment, or may be a result of insufficient spatial resolution to cleanly resolve the magnetic boundary of MFCs. It is unlikely that they indicate convective collapse, since the Stokes $V$ profiles of these pixels do not have the characteristic third lobe reported by \citet{nagata2008}. Downflows at the edges of magnetic features were previously found by \citet{vandervoort2005} and \citet{langangen2007}. They are also compatible with the downflows inferred from the modelling of Stokes $V$ asymmetries \citep{grossmann1988,solanki1989,buente1993}. These downflows are strongest at $\log(\tau)=0$, gradually weaken higher up in the atmosphere and have completely disappeared at $\log(\tau)=-2.3$, suggesting that they are associated with the granular convection pattern. However, downflows exceeding 6 km/s as well as supersonic velocities are only found at the boundaries of MFCs and never in the quiet Sun, indicating that the granular downflows are strengthened by the presence of the MFC. We may be seeing a similar effect as at the edges of light-bridge granules, which display extremely strong downflows, probably because of a combination of radiative losses into the magnetic feature, which leads to faster flows, and because the Wilson depression allows us to see deeper layers with faster flows \citep{lagg2014}. The downflows around MFCs also appear qualitatively similar to the downflows observed around pores by \citet{hirzberger2003,sankara2003} and \citet{cho2010} and, therefore, might show a high temporal variation as well.\\ 
Several MFCs also possess a plume-like downflow features characterised by localised strong downflows traceable through all $\log(\tau)$ layers. These plumes become larger in size and smaller in magnitude with height. They also trace the expansion of their MFC by shifting outwards with height. The origin of these plumes might be tied to a particular active granule in the vicinity, or may be caused by plasma pouring down from the chromosphere. A reconnection event in the chromosphere might also give rise to such a structure, but the rapid increase of LOS velocities with depth appears to be incompatible with such a scenario.\\ 
Located within the downflow rings surrounding the MFCs, we found small magnetic patches bearing the opposite polarity to the main MFC to which they adjoin. These opposite polarity patches are only visible in the lower two $\log(\tau)$ nodes, as shown in Fig. \ref{imgIncl}, supporting the notion that they are intimately connected to the downflows in which they are immersed. MHD simulations carried out by \citet{steiner1998} and \citet{voegler2005} predict such patches in the vicinity of strong magnetic field concentrations,  \citep[see Fig. 5 in][]{voegler2005}. Indirect evidence was also found in a network patch by \citet{zayer1989}, although the exact spatial proximity to the MFC could not be established in that publication due to the low spatial resolution. However, the reversal of the Stokes $V$ amplitude in the wings of the 630 nm line pair is only seen in the deconvolved profiles produced by the inversion, where the Stokes $V$ spectrum displays three lobes. An additional complication is the masking of these magnetic fields by the canopy of the main MFC. The canopy has a typical field strength of 300 G or higher at locations overlying such weak opposite polarity fields. The canopy thus produces a strong signal in Stokes $Q$, $U$ and $V$, whereas the field strengths of the opposite polarity patches at $\log(\tau)=0$ are typically well below $\emph{canopy}$ values. This also prevents the selection of these patches via a typical amplitude threshold in the Stokes profile, so that we cannot completely rule out that even those opposite polarity patches with $B$>100 G are an artifact of the inversion. The opposite polarities typically carry less than $1\%$ of the flux present in their parent MFCs. Further observations performed with a higher spatial resolution are necessary to ascertain the existence of these small opposite polarity patches. Such observations would be particularly useful in the 1.56 $\mu$m lines, due to their large Zeeman sensitivity and low formation height. The 1.56 $\mu$m data analysed by \citet{zayer1989}, although of low spatial resolution, provide some support for our results. The close coexistence of such opposite polarities might give rise to current sheets and could lead to reconnection events.\\ 
We have introduced a novel method for the resolution of the $180^\circ$ azimuth ambiguity applicable to largely unipolar regions. It makes use of the basic thin flux-tube structure of MFCs and assumes that the divergence of the magnetic field tends towards zero. With this method we find that the $\emph{core}$ pixels of MFCs have typical inclinations, relative to local solar coordinates, between $10^\circ$ and $15^\circ$ in all three $\log(\tau)$ nodes of the inversion, which agrees well with earlier inclination results found by \citet{topka1992,bernasconi1995} and \citet{martinez1997}. The distribution of the azimuths shows a preference for the eastern direction, which can be attributed to a LOS effect. MFCs located closest to the disc centre in the field of view showed the most homogeneous azimuth distribution, which supports the conclusion of \citet{martinez1997} that in general MFCs have no preferred orientation. MFCs close to the sunspot did, however, display azimuth distributions that were either predominantly directed towards or away from the spot depending on their polarity. The canopies of these MFCs were also irregular and elongated. This demonstrates that a nearby sunspot has a direct impact on the properties of MFCs even in the middle photosphere. \\
The inversion allowed us to clearly differentiate between magnetic fields which form magnetic canopies from those which form the core or root of a MFC. A separate canopy could be identified for all MFCs as well as for the pores and sunspot. The canopies were found to harbour weaker, more horizontal magnetic fields, with inclinations as high as $\Gamma=80^\circ$. The typical LOS velocities in the canopies are identical to the LOS velocities found at the core of a MFC. Many magnetic canopies were found to lie above essentially field free regions. The canopies found here agree with the model proposed by \citet{grossmann1988} for the production of the Stokes $V$ area asymmetry and the observational results presented by \citet{rezaei2007,narayan2010} and \citet{martinezgonzales2012}. All magnetic features were observed to expand with height and many isolated MFCs merged to form comparatively large expanses of magnetic field at $\log(\tau)=-2.3$. We therefore expect that at least some of the magnetic features display a similarity to the wine-glass model described by \citet{buente1993}.\\ 
The average temperature stratification within MFCs most closely followed the empirical plage flux-tube model of \citet{solanki1992pla}. Temperatures were hotter by 300 K within the MFCs when compared to the quiet Sun in the upper two $\log(\tau)$ layers. However, a closer inspection of MFCs revealed that the temperature is not homogeneous in any $\log(\tau)$ across a MFC. The highest temperatures in a MFC are typically positioned at its edges and concentrated into isolated points at $\log(\tau)=0$. Often supersonic downflows were located in the vicinity of these temperature enhancements. This raises the possibility that the two might be connected, since the supersonic flows likely create a shockfront below the $\log(\tau)=0$ layer. However, there were also cases of temperature enhancements without a nearby supersonic velocity. It is likely that a clear picture of the effect of these supersonic downflows upon an MFC can only be determined together with an analysis of a time series of data.\\
The inversion procedure included a depth-dependent microturbulent velocity, with the macroturbulence set to zero. Particularly high values of the microturbulence, partly in excess of 5 km/s at $\log(\tau)=0$, were found at the edges of the magnetic features. The typical microturbulent velocities in MFCs obtained by the inversion has an average value of 2.8 km/s at $\log(\tau)=0$ and becomes weaker with height. Such values are on a par with the broadenings presented by \citet{solanki1986} and \citet{zayer1989}. This suggests the presence of strong unresolved velocities around the MFCs. Also, although its amplitude decreases rapidly with height, $\xi_{mic}$ is at least a factor of 2 larger in the MFC than in the quiet Sun in the upper two $\log(\tau)$ layers. These large microturbulence values may be telling us that the strong downflows at the edges of the MFCs are associated with vigorous turbulent motions, or it may be a signature of waves travelling along the MFC. Finally, the microturbulence returned by the inversion code may be signalling strong horizontal velocity gradients, e.g. across the boundary of MFCs. A possible geometry for such gradients is a strong downflow outside and a weak one inside.\\

\section{Conclusion}
In this investigation we have been able to confirm many previously obtained properties associated with magnetic fields, which form solar plage. At the same time, we have also uncovered new features associated with kG magnetic fields. KiloGauss fields in plage are strong, vertical ($10^\circ - 15^\circ$) magnetic fields on the order of 1.5 kG, which expand with height similar to a thin flux-tube forming extensive canopies as low as the upper photosphere. These fields are further characterised by being on average 300 K hotter than their surroundings in the middle and upper photosphere, in agreement with empirical flux-tube models, and contain weak, on average 200 m/s, plasma flows within them. This investigation has also discovered several new properties of these fields. Each magnetic flux concentration is surrounded by a ring containing strong downflows, on average 2.4 km/s in the lower photosphere. A typical ring shifts outward with height as the field expands and parts of it can attain supersonic velocities in the lower photosphere. Co-spatial with these rings we found enhanced microturbulent velocities decreasing with height as well as magnetic patches situated beneath the canopy with a polarity opposite to the main plage-forming fields. The plasma temperature of the MFCs was not uniform across their cross sections, displaying instead temperature enhancements at their edges. Magnetic field concentrations located close to larger features such as pores or a sunspot displayed an asymmetric canopy and more inclined magnetic fields.

\begin{acknowledgements}
Hinode is a Japanese mission developed and launched by ISAS/JAXA, with NAOJ as domestic partner and NASA and STFC (UK) as international partners. It is operated by these agencies in co-operation with ESA and NSC (Norway).
This work has been partially supported by the BK21 plus program through the National Research Foundation (NRF) funded by the Ministry of Education of Korea.
D.B. acknowledges a PhD fellowship of the International Max 
Planck Research School on Physical Processes in the Solar 
System and Beyond (IMPRS).
\end{acknowledgements}

\bibliographystyle{aa}
\bibliography{/Users/davidbuehler/Documents/Latex/TheBib_copy}{} 

\begin{thebibliography}{106}
\expandafter\ifx\csname natexlab\endcsname\relax\def\natexlab#1{#1}\fi

\bibitem[{{Anstee} \& {O'Mara}(1995)}]{anstee1995}
{Anstee}, S.~D. \& {O'Mara}, B.~J. 1995, \mnras, 276, 859

\bibitem[{{Babcock} \& {Babcock}(1955)}]{babcock1955}
{Babcock}, H.~W. \& {Babcock}, H.~D. 1955, \apj, 121, 349

\bibitem[{{Barklem} \& {O'Mara}(1997)}]{barklem1997}
{Barklem}, P.~S. \& {O'Mara}, B.~J. 1997, \mnras, 290, 102

\bibitem[{{Barklem} {et~al.}(1998){Barklem}, {O'Mara}, \& {Ross}}]{barklem1998}
{Barklem}, P.~S., {O'Mara}, B.~J., \& {Ross}, J.~E. 1998, \mnras, 296, 1057

\bibitem[{{Barthol} {et~al.}(2011){Barthol}, {Gandorfer}, {Solanki},
  {Sch{\"u}ssler}, {Chares}, {Curdt}, {Deutsch}, {Feller}, {Germerott},
  {Grauf}, {Heerlein}, {Hirzberger}, {Kolleck}, {Meller}, {M{\"u}ller},
  {Riethm{\"u}ller}, {Tomasch}, {Kn{\"o}lker}, {Lites}, {Card}, {Elmore},
  {Fox}, {Lecinski}, {Nelson}, {Summers}, {Watt}, {Mart{\'{\i}}nez Pillet},
  {Bonet}, {Schmidt}, {Berkefeld}, {Title}, {Domingo}, {Gasent Blesa}, {Del
  Toro Iniesta}, {L{\'o}pez Jim{\'e}nez}, {{\'A}lvarez-Herrero},
  {Sabau-Graziati}, {Widani}, {Haberler}, {H{\"a}rtel}, {Kampf}, {Levin},
  {P{\'e}rez Grande}, {Sanz-Andr{\'e}s}, \& {Schmidt}}]{barthol2011}
{Barthol}, P., {Gandorfer}, A., {Solanki}, S.~K., {et~al.} 2011, \solphys, 268,
  1

\bibitem[{{Berger} {et~al.}(2004){Berger}, {Rouppe van der Voort},
  {L{\"o}fdahl}, {Carlsson}, {Fossum}, {Hansteen}, {Marthinussen}, {Title}, \&
  {Scharmer}}]{berger2004}
{Berger}, T.~E., {Rouppe van der Voort}, L.~H.~M., {L{\"o}fdahl}, M.~G.,
  {et~al.} 2004, \aap, 428, 613

\bibitem[{{Bernasconi} {et~al.}(1995){Bernasconi}, {Keller}, {Povel}, \&
  {Stenflo}}]{bernasconi1995}
{Bernasconi}, P.~N., {Keller}, C.~U., {Povel}, H.~P., \& {Stenflo}, J.~O. 1995,
  \aap, 302, 533

\bibitem[{{Briand} \& {Solanki}(1998)}]{briand1998}
{Briand}, C. \& {Solanki}, S.~K. 1998, \aap, 330, 1160

\bibitem[{{Bruls} \& {Solanki}(1995)}]{bruls1995}
{Bruls}, J.~H.~M.~J. \& {Solanki}, S.~K. 1995, \aap, 293, 240

\bibitem[{{B\"unte} {et~al.}(1993){B\"unte}, {Solanki}, \&
  {Steiner}}]{buente1993}
{B\"unte}, M., {Solanki}, S.~K., \& {Steiner}, O. 1993, \aap, 268, 736

\bibitem[{{Cho} {et~al.}(2010){Cho}, {Bong}, {Chae}, {Kim}, \&
  {Park}}]{cho2010}
{Cho}, K.-S., {Bong}, S.-C., {Chae}, J., {Kim}, Y.-H., \& {Park}, Y.-D. 2010,
  \apj, 723, 440

\bibitem[{{Danilovic} {et~al.}(2008){Danilovic}, {Gandorfer}, {Lagg},
  {Sch{\"u}ssler}, {Solanki}, {V{\"o}gler}, {Katsukawa}, \&
  {Tsuneta}}]{danilovic2008}
{Danilovic}, S., {Gandorfer}, A., {Lagg}, A., {et~al.} 2008, \aap, 484, L17

\bibitem[{{Danilovic} {et~al.}(2010){Danilovic}, {Sch{\"u}ssler}, \&
  {Solanki}}]{danilovicft2010}
{Danilovic}, S., {Sch{\"u}ssler}, M., \& {Solanki}, S.~K. 2010, \aap, 509, A76

\bibitem[{{Defouw}(1976)}]{defouw1976}
{Defouw}, R.~J. 1976, \apj, 209, 266

\bibitem[{{Deinzer} {et~al.}(1984){Deinzer}, {Hensler}, {Sch\"ussler}, \&
  {Weisshaar}}]{deinzer1984}
{Deinzer}, W., {Hensler}, G., {Sch\"ussler}, M., \& {Weisshaar}, E. 1984, \aap,
  139, 435

\bibitem[{{Elmore} {et~al.}(1992){Elmore}, {Lites}, {Tomczyk}, {Skumanich},
  {Dunn}, {Schuenke}, {Streander}, {Leach}, {Chambellan}, \&
  {Hull}}]{elmore1992}
{Elmore}, D.~F., {Lites}, B.~W., {Tomczyk}, S., {et~al.} 1992, in Society of
  Photo-Optical Instrumentation Engineers (SPIE) Conference Series, Vol. 1746,
  Polarization Analysis and Measurement, ed. D.~H. {Goldstein} \& R.~A.
  {Chipman}, 22--33

\bibitem[{{Frazier} \& {Stenflo}(1972)}]{frazier1972}
{Frazier}, E.~N. \& {Stenflo}, J.~O. 1972, \solphys, 27, 330

\bibitem[{{Frutiger} {et~al.}(2000){Frutiger}, {Solanki}, {Fligge}, \&
  {Bruls}}]{frutiger2000}
{Frutiger}, C., {Solanki}, S.~K., {Fligge}, M., \& {Bruls}, J.~H.~M.~J. 2000,
  \aap, 358, 1109

\bibitem[{{Giovanelli}(1982)}]{giovanelli1982}
{Giovanelli}, R.~G. 1982, \solphys, 80, 21

\bibitem[{{Grossmann-Doerth} {et~al.}(1988){Grossmann-Doerth}, {Sch\"ussler},
  \& {Solanki}}]{grossmann1988}
{Grossmann-Doerth}, U., {Sch\"ussler}, M., \& {Solanki}, S.~K. 1988, \aap, 206,
  L37

\bibitem[{{Grossmann-Doerth} {et~al.}(1998){Grossmann-Doerth}, {Sch\"ussler},
  \& {Steiner}}]{grossmann1998}
{Grossmann-Doerth}, U., {Sch\"ussler}, M., \& {Steiner}, O. 1998, \aap, 337,
  928

\bibitem[{{Hale}(1908)}]{hale1908}
{Hale}, G.~E. 1908, \apj, 28, 315

\bibitem[{{Hirzberger}(2003)}]{hirzberger2003}
{Hirzberger}, J. 2003, \aap, 405, 331

\bibitem[{{Holweger} {et~al.}(1978){Holweger}, {Gehlsen}, \&
  {Ruland}}]{holweger1978}
{Holweger}, H., {Gehlsen}, M., \& {Ruland}, F. 1978, \aap, 70, 537

\bibitem[{{Howard} \& {Stenflo}(1972)}]{howard1972}
{Howard}, R. \& {Stenflo}, J.~O. 1972, \solphys, 22, 402

\bibitem[{{Ichimoto} {et~al.}(2008){Ichimoto}, {Lites}, {Elmore}, {Suematsu},
  {Tsuneta}, {Katsukawa}, {Shimizu}, {Shine}, {Tarbell}, {Title}, {Kiyohara},
  {Shinoda}, {Card}, {Lecinski}, {Streander}, {Nakagiri}, {Miyashita},
  {Noguchi}, {Hoffmann}, \& {Cruz}}]{ichimoto2008sot}
{Ichimoto}, K., {Lites}, B., {Elmore}, D., {et~al.} 2008, \solphys, 249, 233

\bibitem[{{Ishikawa} \& {Tsuneta}(2009)}]{ishikawa2009}
{Ishikawa}, R. \& {Tsuneta}, S. 2009, \aap, 495, 607

\bibitem[{{Ishikawa} {et~al.}(2008){Ishikawa}, {Tsuneta}, {Ichimoto}, {Isobe},
  {Katsukawa}, {Lites}, {Nagata}, {Shimizu}, {Shine}, {Suematsu}, {Tarbell}, \&
  {Title}}]{ishikawa2008}
{Ishikawa}, R., {Tsuneta}, S., {Ichimoto}, K., {et~al.} 2008, \aap, 481, L25

\bibitem[{{Keller} {et~al.}(1990){Keller}, {Stenflo}, {Solanki}, {Tarbell}, \&
  {Title}}]{keller1990}
{Keller}, C.~U., {Stenflo}, J.~O., {Solanki}, S.~K., {Tarbell}, T.~D., \&
  {Title}, A.~M. 1990, \aap, 236, 250

\bibitem[{{Kobel} {et~al.}(2011){Kobel}, {Solanki}, \& {Borrero}}]{kobel2011}
{Kobel}, P., {Solanki}, S.~K., \& {Borrero}, J.~M. 2011, \aap, 531, A112

\bibitem[{{Kobel} {et~al.}(2012){Kobel}, {Solanki}, \& {Borrero}}]{kobel2012}
{Kobel}, P., {Solanki}, S.~K., \& {Borrero}, J.~M. 2012, \aap, 542, A96

\bibitem[{{Kosugi} {et~al.}(2007){Kosugi}, {Matsuzaki}, {Sakao}, {Shimizu},
  {Sone}, {Tachikawa}, {Hashimoto}, {Minesugi}, {Ohnishi}, {Yamada}, {Tsuneta},
  {Hara}, {Ichimoto}, {Suematsu}, {Shimojo}, {Watanabe}, {Shimada}, {Davis},
  {Hill}, {Owens}, {Title}, {Culhane}, {Harra}, {Doschek}, \&
  {Golub}}]{kosugi2007}
{Kosugi}, T., {Matsuzaki}, K., {Sakao}, T., {et~al.} 2007, \solphys, 243, 3

\bibitem[{{Kuckein} {et~al.}(2012){Kuckein}, {Mart{\'{\i}}nez Pillet}, \&
  {Centeno}}]{kuckein2012}
{Kuckein}, C., {Mart{\'{\i}}nez Pillet}, V., \& {Centeno}, R. 2012, \aap, 539,
  A131

\bibitem[{{Lagg} {et~al.}(2010){Lagg}, {Solanki}, {Riethm{\"u}ller},
  {Mart{\'{\i}}nez Pillet}, {Sch{\"u}ssler}, {Hirzberger}, {Feller}, {Borrero},
  {Schmidt}, {del Toro Iniesta}, {Bonet}, {Barthol}, {Berkefeld}, {Domingo},
  {Gandorfer}, {Kn{\"o}lker}, \& {Title}}]{lagg2010}
{Lagg}, A., {Solanki}, S.~K., {Riethm{\"u}ller}, T.~L., {et~al.} 2010, \apjl,
  723, L164

\bibitem[{{Lagg} {et~al.}(2014){Lagg}, {Solanki}, {van Noort}, \&
  {Danilovic}}]{lagg2014}
{Lagg}, A., {Solanki}, S.~K., {van Noort}, M., \& {Danilovic}, S. 2014, \aap,
  568, A60

\bibitem[{{Langangen} {et~al.}(2007){Langangen}, {Carlsson}, {Rouppe van der
  Voort}, \& {Stein}}]{langangen2007}
{Langangen}, {\O}., {Carlsson}, M., {Rouppe van der Voort}, L., \& {Stein},
  R.~F. 2007, \apj, 655, 615

\bibitem[{{Lawrence} {et~al.}(1993){Lawrence}, {Topka}, \&
  {Jones}}]{lawrence1993}
{Lawrence}, J.~K., {Topka}, K.~P., \& {Jones}, H.~P. 1993, \jgr, 98, 18911

\bibitem[{{Leka} {et~al.}(2009){Leka}, {Barnes}, {Crouch}, {Metcalf}, {Gary},
  {Jing}, \& {Liu}}]{leka2009}
{Leka}, K.~D., {Barnes}, G., {Crouch}, A.~D., {et~al.} 2009, \solphys, 260, 83

\bibitem[{{Lin}(1995)}]{lin1995}
{Lin}, H. 1995, \apj, 446, 421

\bibitem[{{Lites}(1973)}]{lites1973}
{Lites}, B.~W. 1973, \solphys, 32, 283

\bibitem[{{Lites} \& {Ichimoto}(2013)}]{lites2013}
{Lites}, B.~W. \& {Ichimoto}, K. 2013, \solphys, 283, 601

\bibitem[{{Mart{\'{\i}}nez Gonz{\'a}lez} {et~al.}(2012){Mart{\'{\i}}nez
  Gonz{\'a}lez}, {Bellot Rubio}, {Solanki}, {Mart{\'{\i}}nez Pillet}, {Del Toro
  Iniesta}, {Barthol}, \& {Schmidt}}]{martinezgonzales2012}
{Mart{\'{\i}}nez Gonz{\'a}lez}, M.~J., {Bellot Rubio}, L.~R., {Solanki}, S.~K.,
  {et~al.} 2012, \apjl, 758, L40

\bibitem[{{Mart{\'{\i}}nez Pillet} {et~al.}(2011){Mart{\'{\i}}nez Pillet}, {Del
  Toro Iniesta}, {{\'A}lvarez-Herrero}, {Domingo}, {Bonet}, {Gonz{\'a}lez
  Fern{\'a}ndez}, {L{\'o}pez Jim{\'e}nez}, {Pastor}, {Gasent Blesa}, {Mellado},
  {Piqueras}, {Aparicio}, {Balaguer}, {Ballesteros}, {Belenguer}, {Bellot
  Rubio}, {Berkefeld}, {Collados}, {Deutsch}, {Feller}, {Girela}, {Grauf},
  {Heredero}, {Herranz}, {Jer{\'o}nimo}, {Laguna}, {Meller}, {Men{\'e}ndez},
  {Morales}, {Orozco Su{\'a}rez}, {Ramos}, {Reina}, {Ramos},
  {Rodr{\'{\i}}guez}, {S{\'a}nchez}, {Uribe-Patarroyo}, {Barthol}, {Gandorfer},
  {Knoelker}, {Schmidt}, {Solanki}, \& {Vargas
  Dom{\'{\i}}nguez}}]{martinez2011}
{Mart{\'{\i}}nez Pillet}, V., {Del Toro Iniesta}, J.~C., {{\'A}lvarez-Herrero},
  A., {et~al.} 2011, \solphys, 268, 57

\bibitem[{{Mart{\'{\i}}nez Pillet} {et~al.}(1997){Mart{\'{\i}}nez Pillet},
  {Lites}, \& {Skumanich}}]{martinez1997}
{Mart{\'{\i}}nez Pillet}, V., {Lites}, B.~W., \& {Skumanich}, A. 1997, \apj,
  474, 810

\bibitem[{{Metcalf} {et~al.}(2006){Metcalf}, {Leka}, {Barnes}, {Lites},
  {Georgoulis}, {Pevtsov}, {Balasubramaniam}, {Gary}, {Jing}, {Li}, {Liu},
  {Wang}, {Abramenko}, {Yurchyshyn}, \& {Moon}}]{metcalf2006}
{Metcalf}, T.~R., {Leka}, K.~D., {Barnes}, G., {et~al.} 2006, \solphys, 237,
  267

\bibitem[{{Morinaga} {et~al.}(2008){Morinaga}, {Sakurai}, {Ichimoto},
  {Yokoyama}, {Shimojo}, \& {Katsukawa}}]{moriaga2008}
{Morinaga}, S., {Sakurai}, T., {Ichimoto}, K., {et~al.} 2008, \aap, 481, L29

\bibitem[{{Nagata} {et~al.}(2008){Nagata}, {Tsuneta}, {Suematsu}, {Ichimoto},
  {Katsukawa}, {Shimizu}, {Yokoyama}, {Tarbell}, {Lites}, {Shine}, {Berger},
  {Title}, {Bellot Rubio}, \& {Orozco Su{\'a}rez}}]{nagata2008}
{Nagata}, S., {Tsuneta}, S., {Suematsu}, Y., {et~al.} 2008, \apjl, 677, L145

\bibitem[{{Narayan} \& {Scharmer}(2010)}]{narayan2010}
{Narayan}, G. \& {Scharmer}, G.~B. 2010, \aap, 524, A3

\bibitem[{{Okamoto} {et~al.}(2008){Okamoto}, {Tsuneta}, {Lites}, {Kubo},
  {Yokoyama}, {Berger}, {Ichimoto}, {Katsukawa}, {Nagata}, {Shibata},
  {Shimizu}, {Shine}, {Suematsu}, {Tarbell}, \& {Title}}]{okamoto2008}
{Okamoto}, T.~J., {Tsuneta}, S., {Lites}, B.~W., {et~al.} 2008, \apjl, 673,
  L215

\bibitem[{{Okamoto} {et~al.}(2009){Okamoto}, {Tsuneta}, {Lites}, {Kubo},
  {Yokoyama}, {Berger}, {Ichimoto}, {Katsukawa}, {Nagata}, {Shibata},
  {Shimizu}, {Shine}, {Suematsu}, {Tarbell}, \& {Title}}]{okamoto2009}
{Okamoto}, T.~J., {Tsuneta}, S., {Lites}, B.~W., {et~al.} 2009, \apj, 697, 913

\bibitem[{{Osherovich} {et~al.}(1983){Osherovich}, {Chapman}, \&
  {Fla}}]{osherovich1983}
{Osherovich}, V.~A., {Chapman}, G.~A., \& {Fla}, T. 1983, \apj, 268, 412

\bibitem[{{Parker}(1978)}]{parker1978}
{Parker}, E.~N. 1978, \apj, 221, 368

\bibitem[{{Pietarila} {et~al.}(2010){Pietarila}, {Cameron}, \&
  {Solanki}}]{pietarila2010expan}
{Pietarila}, A., {Cameron}, R., \& {Solanki}, S.~K. 2010, \aap, 518, A50

\bibitem[{{Pneuman} {et~al.}(1986){Pneuman}, {Solanki}, \&
  {Stenflo}}]{pneuman1986}
{Pneuman}, G.~W., {Solanki}, S.~K., \& {Stenflo}, J.~O. 1986, \aap, 154, 231

\bibitem[{{Rabin}(1992)}]{rabin1992}
{Rabin}, D. 1992, \apj, 391, 832

\bibitem[{{Requerey} {et~al.}(2014){Requerey}, {Del Toro Iniesta}, {Bellot
  Rubio}, {Bonet}, {Mart{\'{\i}}nez Pillet}, {Solanki}, \&
  {Schmidt}}]{requerey2014}
{Requerey}, I.~S., {Del Toro Iniesta}, J.~C., {Bellot Rubio}, L.~R., {et~al.}
  2014, \apj, 789, 6

\bibitem[{{Rezaei} {et~al.}(2007){Rezaei}, {Steiner}, {Wedemeyer-B{\"o}hm},
  {Schlichenmaier}, {Schmidt}, \& {Lites}}]{rezaei2007}
{Rezaei}, R., {Steiner}, O., {Wedemeyer-B{\"o}hm}, S., {et~al.} 2007, \aap,
  476, L33

\bibitem[{{Riethm{\"u}ller} {et~al.}(2013){Riethm{\"u}ller}, {Solanki}, {van
  Noort}, \& {Tiwari}}]{riethmueller2013}
{Riethm{\"u}ller}, T.~L., {Solanki}, S.~K., {van Noort}, M., \& {Tiwari}, S.~K.
  2013, \aap, 554, A53

\bibitem[{{Rouppe van der Voort} {et~al.}(2005){Rouppe van der Voort},
  {Hansteen}, {Carlsson}, {Fossum}, {Marthinussen}, {van Noort}, \&
  {Berger}}]{vandervoort2005}
{Rouppe van der Voort}, L.~H.~M., {Hansteen}, V.~H., {Carlsson}, M., {et~al.}
  2005, \aap, 435, 327

\bibitem[{{R{\"u}edi} {et~al.}(1992){R{\"u}edi}, {Solanki}, {Livingston}, \&
  {Stenflo}}]{rueedi1992}
{R{\"u}edi}, I., {Solanki}, S.~K., {Livingston}, W., \& {Stenflo}, J.~O. 1992,
  \aap, 263, 323

\bibitem[{{Sankarasubramanian} \& {Rimmele}(2003)}]{sankara2003}
{Sankarasubramanian}, K. \& {Rimmele}, T. 2003, \apj, 598, 689

\bibitem[{{Sasso} {et~al.}(2011){Sasso}, {Lagg}, \& {Solanki}}]{sasso2011}
{Sasso}, C., {Lagg}, A., \& {Solanki}, S.~K. 2011, \aap, 526, A42

\bibitem[{{Scharmer} {et~al.}(2003){Scharmer}, {Bjelksjo}, {Korhonen},
  {Lindberg}, \& {Petterson}}]{scharmer2003}
{Scharmer}, G.~B., {Bjelksjo}, K., {Korhonen}, T.~K., {Lindberg}, B., \&
  {Petterson}, B. 2003, in Society of Photo-Optical Instrumentation Engineers
  (SPIE) Conference Series, Vol. 4853, Innovative Telescopes and
  Instrumentation for Solar Astrophysics, ed. S.~L. {Keil} \& S.~V. {Avakyan},
  341--350

\bibitem[{{Sch{\"u}ssler}(1986)}]{schuessler1986}
{Sch{\"u}ssler}, M. 1986, in Small Scale Magnetic Flux Concentrations in the
  Solar Photosphere, ed. W.~{Deinzer}, M.~{Kn{\"o}lker}, \& H.~H. {Voigt}, 103

\bibitem[{{Shimizu} {et~al.}(2008){Shimizu}, {Nagata}, {Tsuneta}, {Tarbell},
  {Edwards}, {Shine}, {Hoffmann}, {Thomas}, {Sour}, {Rehse}, {Ito},
  {Kashiwagi}, {Tabata}, {Kodeki}, {Nagase}, {Matsuzaki}, {Kobayashi},
  {Ichimoto}, \& {Suematsu}}]{shimizu2008sot}
{Shimizu}, T., {Nagata}, S., {Tsuneta}, S., {et~al.} 2008, \solphys, 249, 221

\bibitem[{{Socas-Navarro}(2011)}]{socas2011}
{Socas-Navarro}, H. 2011, \aap, 529, A37

\bibitem[{{Socas-Navarro} \& {Manso Sainz}(2005)}]{socas2005}
{Socas-Navarro}, H. \& {Manso Sainz}, R. 2005, \apjl, 620, L71

\bibitem[{{Solanki}(1986)}]{solanki1986}
{Solanki}, S.~K. 1986, \aap, 168, 311

\bibitem[{{Solanki}(1989)}]{solanki1989}
{Solanki}, S.~K. 1989, \aap, 224, 225

\bibitem[{{Solanki}(1993)}]{solanki1993}
{Solanki}, S.~K. 1993, \ssr, 63, 1

\bibitem[{{Solanki} {et~al.}(2010){Solanki}, {Barthol}, {Danilovic}, {Feller},
  {Gandorfer}, {Hirzberger}, {Riethm{\"u}ller}, {Sch{\"u}ssler}, {Bonet},
  {Mart{\'{\i}}nez Pillet}, {del Toro Iniesta}, {Domingo}, {Palacios},
  {Kn{\"o}lker}, {Bello Gonz{\'a}lez}, {Berkefeld}, {Franz}, {Schmidt}, \&
  {Title}}]{solanki2010}
{Solanki}, S.~K., {Barthol}, P., {Danilovic}, S., {et~al.} 2010, \apjl, 723,
  L127

\bibitem[{{Solanki} \& {Brigljevic}(1992)}]{solanki1992pla}
{Solanki}, S.~K. \& {Brigljevic}, V. 1992, \aap, 262, L29

\bibitem[{{Solanki} {et~al.}(1996{\natexlab{a}}){Solanki}, {Finsterle}, \&
  {R{\"u}edi}}]{solanki1996}
{Solanki}, S.~K., {Finsterle}, W., \& {R{\"u}edi}, I. 1996{\natexlab{a}},
  \solphys, 164, 253

\bibitem[{{Solanki} {et~al.}(1999){Solanki}, {Finsterle}, {R{\"u}edi}, \&
  {Livingston}}]{solanki1999}
{Solanki}, S.~K., {Finsterle}, W., {R{\"u}edi}, I., \& {Livingston}, W. 1999,
  \aap, 347, L27

\bibitem[{{Solanki} {et~al.}(1987){Solanki}, {Keller}, \&
  {Stenflo}}]{solanki1987}
{Solanki}, S.~K., {Keller}, C., \& {Stenflo}, J.~O. 1987, \aap, 188, 183

\bibitem[{{Solanki} {et~al.}(1994){Solanki}, {Montavon}, \&
  {Livingston}}]{solanki1994}
{Solanki}, S.~K., {Montavon}, C.~A.~P., \& {Livingston}, W. 1994, \aap, 283,
  221

\bibitem[{{Solanki} {et~al.}(1992){Solanki}, {R\"uedi}, \&
  {Livingston}}]{solanki1992expan}
{Solanki}, S.~K., {R\"uedi}, I., \& {Livingston}, W. 1992, \aap, 263, 339

\bibitem[{{Solanki} \& {Stenflo}(1984)}]{solanki1984}
{Solanki}, S.~K. \& {Stenflo}, J.~O. 1984, \aap, 140, 185

\bibitem[{{Solanki} {et~al.}(1996{\natexlab{b}}){Solanki}, {Zufferey}, {Lin},
  {R\"uedi}, \& {Kuhn}}]{solanki1996in}
{Solanki}, S.~K., {Zufferey}, D., {Lin}, H., {R\"uedi}, I., \& {Kuhn}, J.~R.
  1996{\natexlab{b}}, \aap, 310, L33

\bibitem[{{Spruit}(1976)}]{spruit1976}
{Spruit}, H.~C. 1976, \solphys, 50, 269

\bibitem[{{Spruit}(1979)}]{spruit1979}
{Spruit}, H.~C. 1979, \solphys, 61, 363

\bibitem[{{Steiner} {et~al.}(1998){Steiner}, {Grossmann-Doerth}, {Kn\"olker},
  \& {Sch\"ussler}}]{steiner1998}
{Steiner}, O., {Grossmann-Doerth}, U., {Kn\"olker}, M., \& {Sch\"ussler}, M.
  1998, \apj, 495, 468

\bibitem[{{Steiner} {et~al.}(1996){Steiner}, {Grossmann-Doerth},
  {Sch{\"u}ssler}, \& {Kn{\"o}lker}}]{steiner1996}
{Steiner}, O., {Grossmann-Doerth}, U., {Sch{\"u}ssler}, M., \& {Kn{\"o}lker},
  M. 1996, \solphys, 164, 223

\bibitem[{{Steiner} {et~al.}(1986){Steiner}, {Pneuman}, \&
  {Stenflo}}]{steiner1986}
{Steiner}, O., {Pneuman}, G.~W., \& {Stenflo}, J.~O. 1986, \aap, 170, 126

\bibitem[{{Stenflo}(1973)}]{stenflo1973}
{Stenflo}, J.~O. 1973, \solphys, 32, 41

\bibitem[{{Stenflo}(2010)}]{stenflo2010}
{Stenflo}, J.~O. 2010, \aap, 517, A37

\bibitem[{{Stenflo} \& {Harvey}(1985)}]{stenflo1985}
{Stenflo}, J.~O. \& {Harvey}, J.~W. 1985, \solphys, 95, 99

\bibitem[{{Stenflo} {et~al.}(1984){Stenflo}, {Solanki}, {Harvey}, \&
  {Brault}}]{stenflo1984}
{Stenflo}, J.~O., {Solanki}, S., {Harvey}, J.~W., \& {Brault}, J.~W. 1984,
  \aap, 131, 333

\bibitem[{{Suematsu} {et~al.}(2008){Suematsu}, {Tsuneta}, {Ichimoto},
  {Shimizu}, {Otsubo}, {Katsukawa}, {Nakagiri}, {Noguchi}, {Tamura}, {Kato},
  {Hara}, {Kubo}, {Mikami}, {Saito}, {Matsushita}, {Kawaguchi}, {Nakaoji},
  {Nagae}, {Shimada}, {Takeyama}, \& {Yamamuro}}]{suematsu2008sot}
{Suematsu}, Y., {Tsuneta}, S., {Ichimoto}, K., {et~al.} 2008, \solphys, 249,
  197

\bibitem[{{Title} {et~al.}(1989){Title}, {Tarbell}, {Topka}, {Ferguson},
  {Shine}, \& {SOUP Team}}]{title1989}
{Title}, A.~M., {Tarbell}, T.~D., {Topka}, K.~P., {et~al.} 1989, \apj, 336, 475

\bibitem[{{Tiwari} {et~al.}(2013){Tiwari}, {van Noort}, {Lagg}, \&
  {Solanki}}]{tiwari2013}
{Tiwari}, S.~K., {van Noort}, M., {Lagg}, A., \& {Solanki}, S.~K. 2013, \aap,
  557, A25

\bibitem[{{Topka} {et~al.}(1992){Topka}, {Tarbell}, \& {Title}}]{topka1992}
{Topka}, K.~P., {Tarbell}, T.~D., \& {Title}, A.~M. 1992, \apj, 396, 351

\bibitem[{{Topka} {et~al.}(1997){Topka}, {Tarbell}, \& {Title}}]{topka1997}
{Topka}, K.~P., {Tarbell}, T.~D., \& {Title}, A.~M. 1997, \apj, 484, 479

\bibitem[{{Tsuneta} {et~al.}(2008){Tsuneta}, {Ichimoto}, {Katsukawa}, {Nagata},
  {Otsubo}, {Shimizu}, {Suematsu}, {Nakagiri}, {Noguchi}, {Tarbell}, {Title},
  {Shine}, {Rosenberg}, {Hoffmann}, {Jurcevich}, {Kushner}, {Levay}, {Lites},
  {Elmore}, {Matsushita}, {Kawaguchi}, {Saito}, {Mikami}, {Hill}, \&
  {Owens}}]{tsuneta2008sot}
{Tsuneta}, S., {Ichimoto}, K., {Katsukawa}, Y., {et~al.} 2008, \solphys, 249,
  167

\bibitem[{{Unno}(1956)}]{unno1956}
{Unno}, W. 1956, \pasj, 8, 108

\bibitem[{{van Noort}(2012)}]{vannoort2012}
{van Noort}, M. 2012, \aap, 548, A5

\bibitem[{{van Noort} {et~al.}(2013){van Noort}, {Lagg}, {Tiwari}, \&
  {Solanki}}]{vannoort2013}
{van Noort}, M., {Lagg}, A., {Tiwari}, S.~K., \& {Solanki}, S.~K. 2013, \aap,
  557, A24

\bibitem[{{Venkatakrishnan}(1986)}]{venkatakrishnan1986}
{Venkatakrishnan}, P. 1986, \nat, 322, 156

\bibitem[{{Viticchi{\'e}} \& {S{\'a}nchez Almeida}(2011)}]{viticchie2010}
{Viticchi{\'e}}, B. \& {S{\'a}nchez Almeida}, J. 2011, \aap, 530, A14

\bibitem[{{V{\"o}gler} {et~al.}(2005){V{\"o}gler}, {Shelyag}, {Sch{\"u}ssler},
  {Cattaneo}, {Emonet}, \& {Linde}}]{voegler2005}
{V{\"o}gler}, A., {Shelyag}, S., {Sch{\"u}ssler}, M., {et~al.} 2005, \aap, 429,
  335

\bibitem[{{Wiehr}(1978)}]{wiehr1978}
{Wiehr}, E. 1978, \aap, 69, 279

\bibitem[{{Xu} {et~al.}(2010){Xu}, {Lagg}, \& {Solanki}}]{xu2010}
{Xu}, Z., {Lagg}, A., \& {Solanki}, S.~K. 2010, \aap, 520, A77

\bibitem[{{Yelles Chaouche} {et~al.}(2009){Yelles Chaouche}, {Solanki}, \&
  {Sch{\"u}ssler}}]{yelles2009}
{Yelles Chaouche}, L., {Solanki}, S.~K., \& {Sch{\"u}ssler}, M. 2009, \aap,
  504, 595

\bibitem[{{Zayer} {et~al.}(1989){Zayer}, {Solanki}, \& {Stenflo}}]{zayer1989}
{Zayer}, I., {Solanki}, S.~K., \& {Stenflo}, J.~O. 1989, \aap, 211, 463

\bibitem[{{Zayer} {et~al.}(1990){Zayer}, {Stenflo}, {Keller}, \&
  {Solanki}}]{zayer1990}
{Zayer}, I., {Stenflo}, J.~O., {Keller}, C.~U., \& {Solanki}, S.~K. 1990, \aap,
  239, 356

\bibitem[{{Zirin} \& {Popp}(1989)}]{zirin1989}
{Zirin}, H. \& {Popp}, B. 1989, \apj, 340, 571

\end{thebibliography}

\end{document}